\documentclass[12pt]{article}
\pdfoutput=1

\usepackage{pgfplots}
\usetikzlibrary{decorations.pathmorphing}

\tikzset{snake it/.style={decorate, decoration=snake}}
\pgfplotsset{compat=1.10}
\usepgfplotslibrary{fillbetween}
\usetikzlibrary{patterns}

\usepackage{draft} 
\usepackage{hyperref}
\usepackage{graphicx,color,subfig}
\usepackage{cite}
\usepackage{mciteplus}
\usepackage{skak}
\usepackage{empheq}
\usepackage{tikz}
\usepackage{booktabs}
\usepackage{siunitx}
\usepackage{bbm}
\usepackage{makecell}
\usepackage{float}

\usepackage{bm}
\usepackage{braket}

\DeclareFontFamily{OT1}{pzc}{}
\DeclareFontShape{OT1}{pzc}{m}{it}{<-> s * [1.10] pzcmi7t}{}
\DeclareMathAlphabet{\mathpzc}{OT1}{pzc}{m}{it}

\usetikzlibrary{snakes}
\usetikzlibrary{arrows}
\usetikzlibrary{decorations.pathmorphing}
\tikzset{snake it/.style={decorate, decoration=snake}}
\usetikzlibrary{shapes.misc}
\tikzset{cross/.style={cross out, draw=black, minimum size=2*(#1-\pgflinewidth), inner sep=0pt, outer sep=0pt},
cross/.default={1pt}}

\def\be#1\ee{\begin{align}#1\end{align}}

\makeatletter
\newcommand{\bdryno}{\mathpalette\bdry@no\relax}
\newcommand{\bdry@no}[2]{%
  \mspace{1mu}%
  \vbox{%
    \hbox{$\m@th#1\scriptstyle{\ast}$}
    \nointerlineskip
    \kern.25ex
    \hbox{$\m@th#1\scriptstyle{\ast}$}
    \kern-.06ex
  }%
  \mspace{1mu}%
}
\makeatother

\usetikzlibrary{shapes.misc}

\usetikzlibrary{decorations.markings}
\usetikzlibrary{shapes.misc}

\tikzset{cross/.style={cross out, draw=black, minimum size=2*(#1-\pgflinewidth), inner sep=0pt, outer sep=0pt},
cross/.default={1pt}}

\begin{document}

\unitlength = .8mm

\begin{titlepage}

\begin{center}

\hfill \\
\hfill \\
\vskip 1cm

\title{Exact quantization and analytic continuation}

\author{Barak Gabai, Xi Yin}

\address{
Jefferson Physical Laboratory, Harvard University, \\
Cambridge, MA 02138 USA
}

\email{bgabai@g.harvard.edu, xiyin@g.harvard.edu}

\end{center}

\abstract{
In this paper we give a streamlined derivation of the exact quantization condition (EQC) on the quantum periods of the Schr\"odinger problem in one dimension with a general polynomial potential, based on Wronskian relations. We further generalize the EQC to potentials with a regular singularity, describing spherical symmetric quantum mechanical systems in a given angular momentum sector. We show that the thermodynamic Bethe ansatz (TBA) equations that govern the quantum periods undergo nontrivial monodromies as the angular momentum is analytically continued between integer values in the complex plane. The TBA equations together with the EQC are checked numerically against Hamiltonian truncation at real angular momenta and couplings, and are used to explore the analytic continuation of the spectrum on the complex angular momentum plane in examples.
}
\vfill

\end{titlepage}

\eject

\begingroup
\hypersetup{linkcolor=black}

\tableofcontents

\endgroup

\section{Introduction} 

The analytic continuation of a quantum mechanical system in its parameters has been shown to possess extraordinarily rich structures that encode profound physics \cite{Regge:1959mz, Bender:1969si, Fonseca:2001dc}. A detailed exploration of such analytic structures typically goes beyond the reach of either perturbation theory or non-perturbative numerical methods such as Hamiltonian truncation. It would thus be desirable to analyze such analytic continuation in sufficiently nontrivial integrable models, from which useful lessons may be drawn.

In recent years, starting with the discovery of the ODE/IM correspondence \cite{Dorey:1998pt,Bazhanov:1998wj,Dorey:1999uk,Dorey:2007zx}, it became widely known that a large class of Schr\"odinger problems in one dimensions are integrable, specifically those with polynomial potentials \cite{Balian:1978et, AIHPA_1983__39_3_211_0, DelErHerv_ExactSemi, AIHPA_1999__71_1_1_0, Voros:1999bz, Ito:2018eon} as well as potentials with a regular singularity \cite{Ito:2020ueb}. The basic idea is that one defines the so-called quantum periods, extending the notion of classical Bohr-Sommerfeld periods as a function of the energy and coupling parameters of the system, that can be shown to obey a set of $Y$-system equations, or equivalently, a set of thermodynamical Bethe ansatz (TBA) equations, where $\hbar$ plays the role of the spectral parameter. The TBA equations may be derived either by viewing the quantum period as the Borel resummation of a power series in $\hbar$ that corrects the classical period, or through its algebraic relation with Wronskians between different solutions to the Schr\"odinger equation with prescribed asymptotic behavior on the complex $x$(coordinate)-plane. While the former viewpoint is closely tied to the resurgence program and wall-crossing phenomena \cite{Ito:2018eon,Emery:2020qqu}, in this paper we will adopt the latter perspective, as will be reviewed in section \ref{sec:Exact}.

To determine the spectrum of the Schr\"odinger problem, one further needs to impose the so-called exact quantization condition (EQC), which typically takes the form of a transcendental equation on the quantum periods. While the strategy for finding the EQC for the Schr\"odinger problem with a polynomial potential is known \cite{Voros:1999bz, PhysRevLett.55.2523}, to the best of our knowledge, a streamlined derivation of the EQC for a general polynomial potential has not been presented in the literature. In section \ref{sec:EQC}, we present a simple such derivation using Wronskian relations. The resulting spectrum is checked numerically against, and shown to be consistent with, results from Hamiltonian truncation. The TBA+EQC approach however has the advantage that it allows for the determination of the (analytically continued) spectrum at complex coupling parameters, at high numerical precision.

In a somewhat nontrivial manner, both the TBA and EQC can be generalized to Schr\"odinger problems with a regular singularity, namely the potential $V(x)$ consists of polynomials in $x$ as well as $x^{-1}$ and $x^{-2}$ terms, that may be equivalently viewed as that of a spherically symmetric system in a sector of given angular momentum $\ell$ (which may be analytically continued to a complex value). The TBA equations for such systems have been formulated in \cite{Ito:2020ueb}. However, we will see that the equations of \cite{Ito:2020ueb} are strictly valid only for $-1\leq\ell\leq 0$, and undergoes nontrivial monodromy when continued to real values of $\ell$ outside of $[-1,0]$. This will be explained in section \ref{sec:NotwallCross}. The monodromy-transformed TBA equation, along with the derivation of EQC for Schr\"odingr systems with a regular singularity (section \ref{sec:eqcsing}), are the main results of this paper.
We carry out some preliminary investigation of the analytic continuation in angular momentum in section \ref{sec:AnContEll}.

We also explain how scattering phases in Schr\"odinger problems with an asymptotic region can be extracted by a similar quantization condition in section \ref{app:scatAmp}. We conclude with some future prospectives in section \ref{sec:discussion}. Details concerning wall-crossing of the TBA equations and numerical implementation are discussed in the appendices.

\section{TBA Equations for Quantum Periods}
\label{sec:Exact}

In this section we review the derivation of the TBA equations that govern the quantum periods of the Schr\"odinger system in one dimension, based on Wronskian relations between solutions with different asymptotics. We follow the approach of \cite{Ito:2018eon, Emery:2020qqu} in the case of a polynomial potential, and \cite{Ito:2020ueb} in the case of a potential with a regular singularity.

\subsection{Polynomial potential}
\label{sec:ExactPoly}

The Schr\"odinger equation in one dimension for a polynomial potential of degree $r+1$, after a rescaling of the coordinate, can be put in the form
\begin{equation} \label{eq:schr}
	\left(-\hbar^2 \partial_x^2 + x^{r+1} + \sum_{a=1}^{r} u_{a} x^{r-a}\right) \psi(x) = 0\ .
\end{equation}
Note that the energy is absorbed into the coefficient $u_r$, and will be treated on equal footing as the other parameters of the potential. Without imposing normalizability, there are always two linearly independent solutions for the wave function, each of which can be analytically continued to the entire complex $x$-plane, as well as analytically continued in $u_a$ and $\hbar$.

The equation \ref{eq:schr} is invariant under the so-called Symanzik rotation,
\begin{equation} \label{eq:symanzyk}
	(x,u_a) \mapsto (\omega x, \omega^{a+1} u_a) , ~~~ \omega \equiv e^{\frac{2\pi i }{r+3}},
\end{equation}
which cyclically permutes the wedges ${\cal S}_k$ of the complex plane, defined as
\begin{equation}
\label{eq:sectors}
	\mathcal S_k = \Bigg\{ x \in \mathbb{C}  : \left|\arg(x)-\frac{2\pi k}{r+3}\right| < \frac{\pi}{r+3}\Bigg\}
\end{equation}
for integer $k$.

For given $u_a$ and real positive $\hbar$, let $y(x,u_a,\hbar)$ be a solution to \ref{eq:schr} that decays in ${\cal S}_0$ as $x\to\infty$. Such a solution is unique up to its overall normalization, to be specified below. We will analytically continue $y(x,u_a,\hbar)$ in $\hbar$ by rotating the phase of $\hbar$. As there is an essential singularity at $\hbar=0$, and there can be nontrivial monodromies around $\hbar=0$, we must keep track of the path of the analytic continuation. Henceforth we will write $y(x,u_a, e^{i\A}\hbar)$ for the analytic continuation of $y(x,u_a,\hbar)$ by rotating the phase of $\hbar$ continuously from 0 to $\alpha$.

By applying the Symanzik rotation to $y$, in the sense of analytic continuation, we obtain solutions to the same Schr\"odinger equation but with different asymptotic behaviors,
\ie
\label{ykdef}
	y_k(x,u_a,\hbar) \equiv \omega^{k\over 2} y(\omega^{-k} x, \omega^{-(a+1)k}u_a,\hbar) = \omega^{k\over 2} y(x, u_a,e^{i\pi k}\hbar),
\fe
where the second equality of (\ref{ykdef}) follows from the invariance of the Schr\"odinger equation under simultaneous phase rotations of $x$, $u_a$, and $\hbar$. For real positive $\hbar$, $y_k$ decays in the wedge ${\cal S}_k$ as $x\to \infty$, with the asymptotic form
\ie
\label{eq:asymform}
	y_k(x,u_a,\hbar) \approx {\omega^{k\over 2}\over \sqrt{2i}}\left(\hbar^{-{2\over r+3}}\omega^{-k} x\right)^{n^{(k)}_r} \exp\left[ -\frac{2}{(r+3)\hbar}e^{-i\pi k} x^{\frac{r+3}{2}} \left( 1 + \sum_{m=1}^{\lfloor {r+1\over 2} \rfloor} B_m \hbar^{2m\over r+3} x^{-m} \right) \right] ,
\fe
where $B_m$ are defined by
\begin{equation}
	\left(1+\sum_{a=1}^{r} u_a x^{-1-a}\right)^{1\over 2} \equiv 1+\sum_{m=1}^{\infty} B_m \hbar^{2m\over r+3} x^{-m},
\end{equation}
and the exponent $n_r^{(k)}$ is given by
\begin{equation}
	n_r^{(k)} =
	\left\{
	\begin{array}{ll}
	-\frac{r+1}{4},  & \quad r ~\text{even} \\
	-\frac{r+1}{4}-(-)^{k}B_{\frac{r+3}{2}}, & \quad r ~\text{odd}
	\end{array}
	\right. 
\end{equation}
Note that $y_{k+r+3} = -y_k$, as there is no nontrivial monodromy under the full $2\pi$ Symanzik rotation. 
This is in contrast to the case of a potential with regular singularity, to be discussed in the next section.

The Wronskian between $y_{k_1}$ and $y_{k_2}$ is written as
\ie
\label{wronsk}
	W_{k_1,k_2}(u_a,\hbar) \equiv \hbar^{2\over r+3} 
	\big[ y_{k_1}(x,u_a,\hbar)\partial_x y_{k_2}(x,u_a,\hbar) - y_{k_2}(x,u_a,\hbar)\partial_x y_{k_1}(x,u_a,\hbar) \big].
\fe
As is well known, $W_{k_1,k_2}$ is independent of $x$, and satisfies the Plucker relations,
\begin{equation}
\label{pluckerr}
	W_{k_3,k_4}(\hbar)W_{k_1,k_2}(\hbar) = -W_{k_3,k_2}(\hbar)W_{k_4,k_1}(\hbar)-W_{k_3,k_1}(\hbar)W_{k_2,k_4}(\hbar) .
\end{equation}
Now, we define the Y-functions,
\ie
\label{yfuncs}
	& Y_{2j}(u_a, \hbar) \equiv \frac{W_{-j,j}W_{-j-1,j+1}}{W_{-j-1,-j}W_{j,j+1}}(u_a, \hbar)  ,\\
	& Y_{2j+1}(u_a, \hbar) \equiv \frac{W_{-j-1,j}W_{-j-2,j+1}}{W_{-j-2,-j-1}W_{j,j+1}}(u_a, e^{\pi i\over 2}\hbar)  ,
\fe
where the RHS of the second line is defined again by analytic continuation in the phase of $\hbar$. Note the special cases,
\ie
	Y_0 = Y_{r+1} = 0.
\fe
Using the identity
\ie\label{wkshfone}
	W_{k_1,k_2}(u_a, e^{i\pi}\hbar) = W_{k_1+1,k_2+1}(u_a, \hbar)
\fe
and applying the Plucker relations, one can verify that (\ref{yfuncs}) obey the equations of the Y-system,
\ie
\label{ysystem}
	Y_s(u_a, e^{-i\pi/2}\hbar) Y_s(u_a, e^{i\pi/2}\hbar) = (1+Y_{s+1}(u_a, \hbar))(1+Y_{s-1}(u_a, \hbar)) ,
\fe
where $s=1,2,\cdots,r$.

To solve (\ref{ysystem}) one needs to know the asymptotic behavior of the Y-functions in the semi-classical limit $\hbar\to 0$. Denote by ${\cal S}_k^{\gg}$ the region of ${\cal S}_k$ with sufficiently large $|x|$. Writing $V(x) = x^{r+1} + \sum_{a=1}^r u_a x^{r-a}$, we can write the WKB approximation of $y_k$ as
\ie
\label{ykwkb}
y_k(x,u_a,\hbar) \sim (V(x))^{-{1\over 4}} \exp\left[ -{\delta_k \over \hbar} \int_{x_k}^x \sqrt{V(y)} dy \right]
\fe
for $x\in {\cal S}_k^{\gg}$ and positive real $\hbar$, with a suitable choice of the basepoint $x_k \in {\cal S}_k$. The sign $\delta_k=\pm 1$ is chosen such that $y_k$ decays as $x\to \infty$ in ${\cal S}_k$. This approximation of $y_k$ continues to be valid under analytic continuation in $x$ along a path $x=\gamma(t)$, $t\geq 0$, starting from $\gamma(0)=\widetilde x_k\in {\cal S}_k^{\gg}$, provided that $\delta_k {\rm Re} \left[ \sqrt{V(\gamma(t))}\, \gamma'(t) \right] < 0$ along the path.\footnote{This is a consequence of Vitali's theorem which states that a sequence of locally uniformly bounded holomorphic functions converges to a holomorphic function. It implies that $y_k(x,u_a,\hbar)/y_k^{\rm WKB}(x,u_a,\hbar)$ converges to 1 in the $\hbar\to 0$ limit along an ascending path.} We will refer to the latter as an {\it ascending path}. An example of an ascending path is obtained by taking a solution to $\gamma'(t) = -\delta_k \overline{\sqrt{V(\gamma(t))}}$, or by joining the solutions that end and begin at zeroes of $V(x)$.

Obviously, there are ascending paths from $S_k^{\gg}$ to $S_{k\pm 1}^{\gg}$. This allows us to compute the Wronskian $W_{k,k+1}$ by comparing the asymptotic wave function on the ray between ${\cal S}_k$ and ${\cal S}_{k+1}$, giving the result 
\ie\label{wneigh}
W_{k,k+1}(\hbar) = 	\left\{
	\begin{array}{ll}
	1,  & \quad r ~\text{even}, \\
	\omega^{(-)^{k+1}B_{\frac{r+3}{2}}}, & \quad r ~\text{odd}.
	\end{array}
	\right.
\fe
A particular consequence is that there is an algebraic identity expressing one of the $Y_s$ functions with odd $s$ in terms of the rest.

\begin{figure}[h!]
	\centering
	\includegraphics[width=0.42\linewidth]{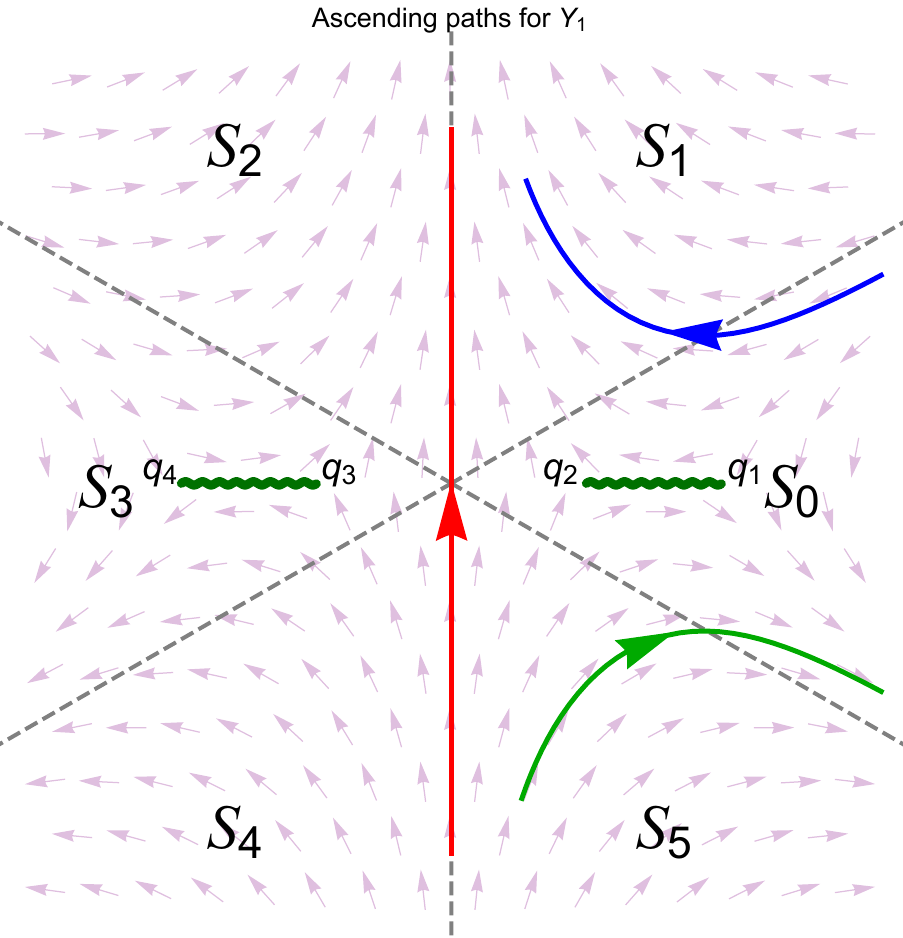}~~~
	\includegraphics[width=0.42\linewidth]{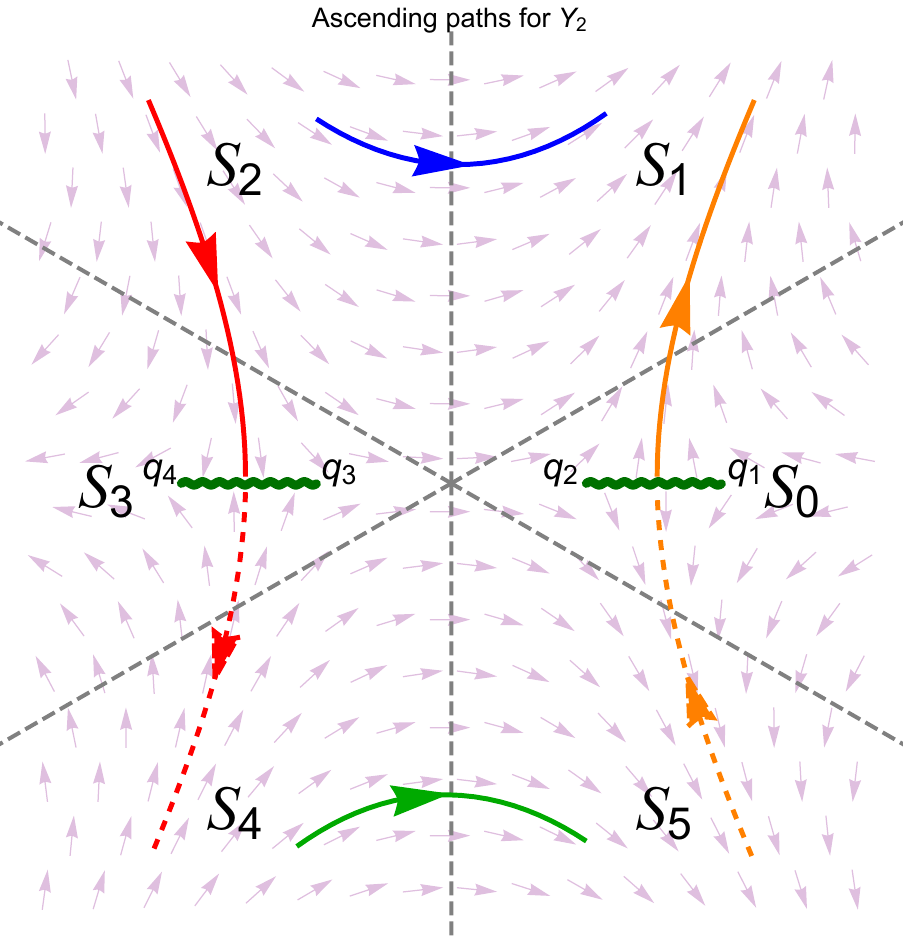}
	\caption{Collections of ascending paths on the complex $x$-plane that determine the semi-classical limit of $Y_1$ (left) and $Y_2$ (right) respectively, in the $r=3$ case (quartic potential).}
	\label{fig:asc}
\end{figure}

More nontrivially, there may also exist ascending paths connecting $S_k^{\gg}$ to $S_{\ell}^{\gg}$ for certain $\ell$'s that differ from $k$ by an odd integer. Let us begin with the special case where all roots of $V(x)=0$, denoted $q_1,\cdots, q_{r+1}$, are real and distinct, ordered according to 
\ie\label{minorder}
q_1 > q_2>\cdots >q_{r+1}.
\fe 
For $j$ in the range $1\leq j\leq \lfloor {r\over 2} \rfloor$, there are ascending paths connecting $S_{j+1}^{\gg}$ to $S_{-j-1}^{\gg}$ that cross the real $x$-axis between $q_{2j+2}$ and $q_{2j+1}$, and ascending paths connecting $S_{-j}^{\gg}$ to $S_{j}^{\gg}$ that cross the real $x$-axis between $q_{2j}$ and $q_{2j-1}$ (see figure \ref{fig:asc}).
This allows us to evaluate the cross ratio of Wronskians that appear in $Y_{2j}$ (\ref{yfuncs}), in the WKB approximation, as
\ie\label{asympeven}
Y_{2j}(u_a, \hbar) \approx e^{ - {m_{2j}\over \hbar} },~~~~m_{2j}\equiv \oint_{\gamma_{2j}} \sqrt{V(x)} dx ,
\fe
where $\gamma_{2j}$ is a counterclockwise contour that encloses $[q_{2j}, q_{2j+1}]$. In a similar way, one can show that $Y_{2j+1}$ is approximated in the semi-classical limit by
\ie\label{asympodd}
Y_{2j+1}(u_a, \hbar) \approx e^{ - {m_{2j+1}\over \hbar} },~~~~ m_{2j+1}\equiv  i \oint_{\gamma_{2j+1}} \sqrt{V(x)} dx,
\fe
where $\gamma_{2j+1}$ is  a counterclockwise contour that encloses $[q_{2j}, q_{2j-1}]$.
The choice of branch here is such that when (\ref{minorder}) is obeyed, $m_s$ is positive for all $s=1,\cdots, r$.

We now define the spectral parameter $\theta$ as
\ie
\hbar = e^{-\theta},
\fe 
and write
\ie\label{yepsrel}
Y_s(u_a, \hbar) \equiv \exp\left[ -\epsilon_s(u_a, \theta) \right].
\fe
We will omit writing the arguments $u_a$ below. As a function of $\theta$, $Y_s$ is expected to be meromorphic on the complex $\theta$-plane with possible poles and zeroes when some of the Wronskians vanish. An assumption we shall now make, which will be justified a posteriori, is that when (\ref{minorder}) is obeyed, $Y_s$ has no zeroes or poles in the strip $-{\pi\over 2} < {\rm Im}\theta < {\pi\over 2}$. In this case, we can rewrite $f_s(\theta) \equiv \epsilon_s(\theta) -  m_s e^\theta$ as the contour integral
\ie{}
f_s(\theta) &= \left(\int_{\mathbb{R}-{i\pi\over 2}} - \int_{\mathbb{R}+{i\pi\over 2}} \right) {d\theta'\over 2\pi i} {f_s(\theta')\over \sinh(\theta'-\theta)}
\\
&= - \int_{-\infty}^\infty {d\theta'\over 2\pi} {f_s(\theta' - {i\pi\over 2}) + f_s(\theta' + {\pi\over 2}) \over \cosh(\theta'-\theta)}
=  - \int_{-\infty}^\infty {d\theta'\over 2\pi} {\epsilon_s(\theta' - {i\pi\over 2}) + \epsilon_s(\theta' + {\pi\over 2}) \over \cosh(\theta'-\theta)}.
\fe
Applying the Y-system equation (\ref{ysystem}) to the integrand on the RHS, we arrive at the TBA equations for $\epsilon_s(\theta)$,
\ie\label{tbaeqns}
\epsilon_s( \theta) = m_s e^\theta - K\star L_{s+1}(\theta) - K\star L_{s-1}(\theta),
\fe
for $s=1,\cdots, r$, where
\ie{}
& K(\theta) \equiv {1\over 2\pi\cosh\theta},
\\
& L_s(\theta)\equiv \log (1+e^{-\epsilon_s(\theta)}),~~~1\leq s\leq r, ~~~~L_0 \equiv L_{r+1}\equiv 0,
\fe
and the convolution $\star$ is defined as $f\star g (\theta)\equiv \int \frac{d \theta'}{2\pi} f(\theta-\theta')g(\theta')$. The advantage of the TBA equations is that they explicitly incorporate the $\hbar\to 0$ asymptotics (\ref{asympeven}), (\ref{asympodd}), and can be solved numerically by iteration, starting with the semi-classical approximation $\epsilon_s^{(0)}(\theta) = m_s e^\theta$.

As we vary the parameters of the potential, the turning points may move away from the real axis. In this process, the collection of ascending paths that determine the semi-classical limit of Wronskians of non-adjacent sectors may change discontinuously, leading to a jump in the semi-classical limit of the Wronskians in question, a phenomenon known as wall-crossing. After wall-crossing, while the Y-system remain as described, the semi-classical limits of the Y-functions jump, leading to a different set of TBA equations. In the literature \cite{Emery:2020qqu,Ito:2020ueb,Ito:2018eon} the different domains of parameter space separated by wall-crossing are often referred to as chambers, the chamber containing (\ref{minorder}) dubbed the ``minimal chamber". An efficient way of deriving the TBA equations in the non-minimal chambers is through inspecting an analytic continuation of (\ref{tbaeqns}) (see \cite{Ito:2018eon,Emery:2020qqu,Alday:2010vh} or appendix \ref{sec:wallCrossing}).

\begin{figure}[h!]
	\centering
	\includegraphics[width=0.42\linewidth]{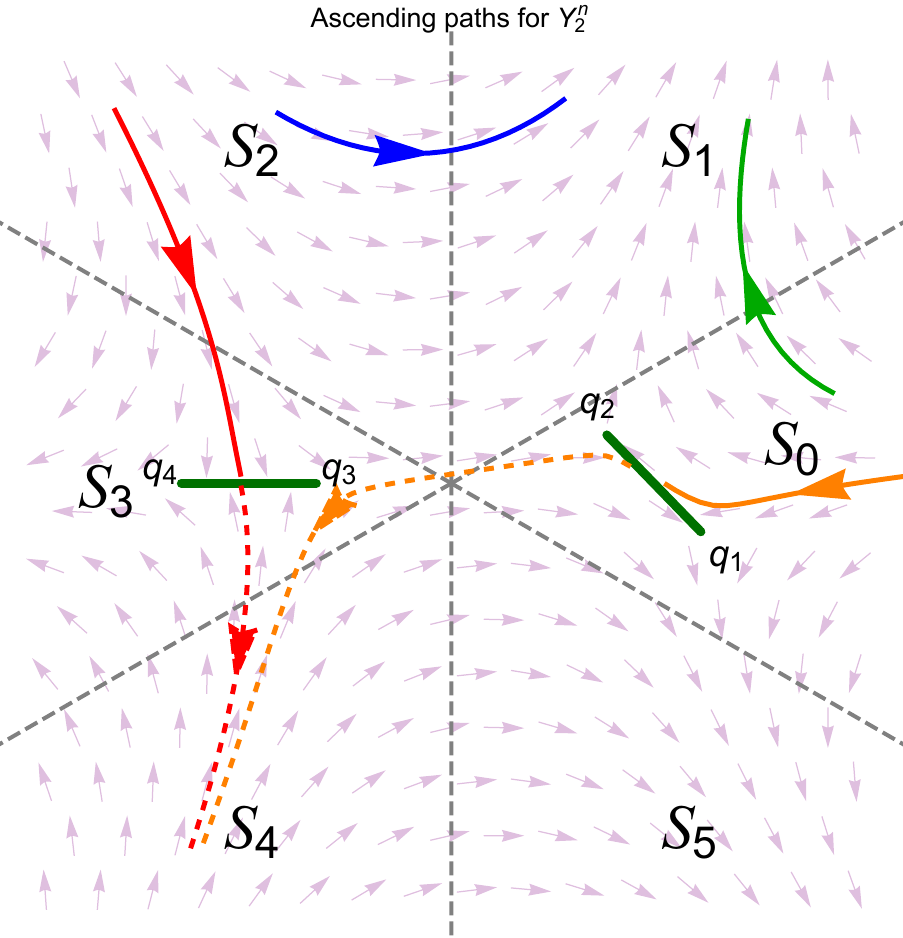}~~~
	\includegraphics[width=0.42\linewidth]{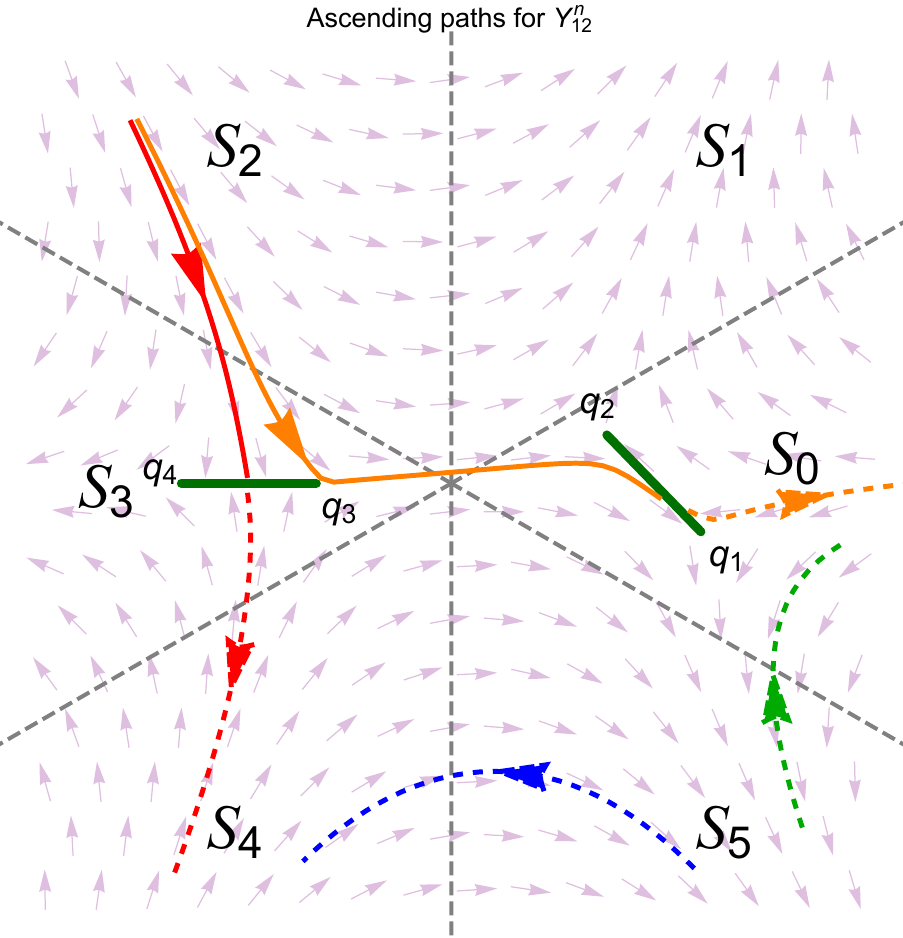}
	\caption[]{Collections of ascending paths for the analytically continued quartic potential $V(x) = \frac{1}{4}(x+2)(x+1)\left(2x-3-e^{-i\phi}\right)\left(2x-3+e^{-i\phi}\right)$ at $\phi=0.8$ (which resides in a non-minimal chamber) that determine the Y-functions $Y_1^n$ (left) and $Y_{12}^n$  (right) respectively. 
	}
	\label{fig:wallcrossedAcsents}
\end{figure}

Let us illustrate how the wall-crossing occurs from the perspective of ascending paths, in an example where the potential is taken to be $V(x) = \frac{1}{4}(x+2)(x+1)\left(2x-3-e^{-i\phi}\right)\left(2x-3+e^{-i\phi}\right)$. If we take $\phi=0$, the roots satisfy (\ref{minorder}) and we are in the minimal chamber. As we increase $\phi$, wall-crossing occurs at $\phi \approx 0.689$. The collections of ascending paths that determine the semi-classical limit of suitable cross ratios of Wronskians after the wall-crossing are shown in figure \ref{fig:wallcrossedAcsents} (compared with figure  \ref{fig:asc} in the minimal chamber). The new Y-functions with simple semi-classical limits after the wall-crossing may be chosen as
\ie
&		Y^n_1(e^{i\phi_1} \hbar ) = \frac{W_{-1,0}W_{1,2}}{W_{0,1}W_{-1,2}}(e^{i\left(\phi_1+\frac{i\pi}{2}\right)}\hbar),~~~~
&		Y^n_2(e^{i\phi_2} \hbar) = \frac{W_{0,1}W_{-2,2}}{W_{-2,0}W_{1,2}}(e^{i\phi_2} \hbar), \\
&		Y^n_{12}(e^{i\phi_{12}} \hbar) = \frac{W_{-1,0}W_{-2,2}}{W_{0,2}W_{-2,-1}}(e^{i\left(\phi_{12}+\frac{i\pi}{2}\right)}\hbar).
\fe
Here $\phi_j$ are the arguments of the classical ``masses"; see appendix \ref{sec:wallCrossing} (where the same result is derived from the analytic continuation of the TBA) for definitions and further details.\footnote{While the definitions of the $Y^n$'s given in appendix \ref{sec:wallCrossing} may appear different, they are in fact related by straightforward applications of Plucker identities.}
In particular, the semi-classical limit of $Y^n_{12}(e^{i\phi_{12}} \hbar)$ is governed by the cycle $\gamma_{12}$ which encloses $q_1$ and $q_3$.

\subsection{Potential with a regular singularity}
\label{sec:exact_sing}

Now we investigate the Schr\"odinger system of the form
\begin{equation} \label{eq:schrSing}
	\left(-\hbar^2 \partial_x^2 + x^{r+1} + \sum_{a=0}^{r+1} u_{a} x^{r-a} + \frac{\hbar^2 \ell(\ell+1)}{x^2}\right) \psi(x) = 0 ,
\end{equation}
following the analysis of \cite{Ito:2020ueb}. We will consider the wedge ${\cal S}_k$ on the complex $x$-plane as before, and the solution $y(x,u_a, \hbar)$ which decays in ${\cal S}_0$ at large $x$ for positive real $\hbar$. $y_k(x,u_a,\hbar)$ are still defined by (\ref{ykdef})
where the Symanzik rotation is defined via analytic continuation (leaving $\ell$ invariant). A key difference however is that now the solutions may have a nontrivial monodromy around $x=0$, and consequently $y_k$ is not the same as $y_{k+r+3}$.

We can analytically extend $y_k(x,u_a,\hbar)$ to the complex $x$-plane away from the branch cut $\{x:{\rm arg}(x) = -{\pi\over r+3}\}$ (the lower edge of ${\cal S}_0$), in order to compare different $y_k$'s at the same point. This allows for the Wronskian $W_{k_1, k_2}$ to be defined as in (\ref{wronsk}). In particular, we can express $y_k$ as a linear combination of $y_0$ and $y_1$ via
\ie
y_k = \frac{W_{k,1}}{W_{0,1}}y_0+\frac{W_{0,k}}{W_{0,1}}y_1.
\fe
The Y-functions $Y_s(u_a,\hbar)$ are defined as (\ref{yfuncs}). While we still have $Y_0=0$ by definition, $Y_{r+1}$ is no longer zero due to the monodromy. The equation (\ref{ysystem}) still holds for $s=1,\cdots, r$, with a nontrivial $Y_{r+1}$, but also extends to $s\geq r+1$. The Y-system will be closed by putting an additional constraint on $Y_{r+2}$, which will follow from the consideration of monodromy below.

Specifically, $y_{r+3}$ and $y_{r+4}$ are related to $y_0$ and $y_1$ by the monodromy $x\mapsto e^{-2\pi i} x$. This allows us to constrain their Wronskians via
\ie{}
{y_{r+4} \choose
	y_{r+3} } (x) \equiv \omega^{r+3\over 2}{y_1\choose y_0}(e^{-2\pi i} x) 
= \omega^{{r+3\over 2}} \Omega(u_a,\hbar,\ell){y_1\choose y_0}(x)  \ ,
\fe
where,
\begin{equation}
	\Omega(u_a,\hbar,\ell) \equiv \omega^{-{r+3\over 2}}\begin{pmatrix} \frac{W_{0,r+4}}{W_{0,1}} & \frac{W_{r+4,1}}{W_{0,1}} \\
	\frac{W_{0,r+3}}{W_{0,1}} & \frac{W_{r+3,1}}{W_{0,1}} 
	\end{pmatrix} 
\end{equation}
is the monodromy matrix acting on $(y_0, y_1)$. On the other hand, there is a basis of solutions $(y_+, y_-)$ that behave near $x=0$ as
\ie\label{ynearorign}
	y_{\pm} \sim x^{{1\over 2} \pm (\ell+{1\over 2})}.
\fe
In this basis the monodromy takes the simple form
\ie\label{ypmmono}
{y_+ \choose y_-} (e^{-2\pi i}x) = \begin{pmatrix}
e^{-2\pi i \ell} & 0 \\
0 & e^{2\pi i \ell} 
\end{pmatrix} {y_+ \choose y_- }(x) .
\fe
While we do not know a priori the linear transformation between $(y_0, y_1)$ and $(y_+, y_-)$, (\ref{ypmmono}) determines the eigenvalues of $\Omega(u_a, \hbar, \ell)$. As $\Omega$ is unimodular, the nontrivial constraint is
\ie\label{trmono}
{\rm Tr}\, \Omega(u_a,\hbar,\ell) \equiv \omega^{-{r+3\over 2}}\left(\frac{W_{0,r+4}}{W_{r+3,r+4}}+\frac{W_{r+3,1}}{W_{0,1}}\right) = 2\cos(2 \pi \ell) .
\fe
Let us now introduce
\ie \label{eq:Yhat}
\hat Y(u_a,\hbar,\ell) \equiv \frac{W_{0,r+2}}{W_{r+2,r+3}}(u_a, e^{-i\pi{r+2\over 2}}\hbar,\ell),
\fe
which obeys the Plucker relation 
\ie\label{yhatpluck}
\hat Y(u_a, e^{-{i\pi\over 2}}\hbar,\ell) \hat Y(u_a, e^{i\pi\over 2}\hbar,\ell) &= 1+Y_{r+1}(u_a, \hbar,\ell).
\fe
(\ref{trmono}) can be equivalently rewritten as an algebraic relation between  $\hat Y(u_a,\hbar,\ell)$ and $Y_{r+2}(u_a, \hbar,\ell) = {W_{0,r+2}W_{-1,r+3}\over W_{-1,0}W_{r+2,r+3}}(u_a, e^{-i\pi {r+2\over 2}}\hbar,\ell)$,
\ie\label{ytwohat}
\omega^{-{r+3\over 2}}\left( {Y_{r+2}\over \hat Y} - \hat Y \right) = 2\cos(2\pi \ell)  .
\fe
After using equation (\ref{ytwohat}) to replace $Y_{r+2}$ with $\hat Y$ in (\ref{ysystem}), equation (\ref{yhatpluck}) closes the system of Y-functions $Y_1,\cdots, Y_{r+1}, \hat Y$. The full set of Y-system equations are
\ie
\label{ysystem_sing}
& Y_s(u_a, e^{-{i\pi\over 2}}\hbar,\ell) Y_s(u_a, e^{{i\pi\over 2}}\hbar,\ell) = (1+Y_{s+1}(u_a, \hbar,\ell))(1+Y_{s-1}(u_a, \hbar,\ell)) , ~~~~ s=1,\cdots,r, ~~(Y_0=0)
\\
& Y_{r+1}(u_a, e^{-{i\pi\over 2}}\hbar,\ell) Y_{r+1}(u_a, e^{{i\pi\over 2}}\hbar,\ell) = (1+\omega^{r+3\over 2}e^{2\pi i \ell}\hat Y(u_a, \hbar,\ell))(1+\omega^{r+3\over 2}e^{-2\pi i \ell}\hat Y(u_a, \hbar,\ell))(1+Y_{r}(u_a, \hbar,\ell)),
\fe
together with (\ref{yhatpluck}).

Note that the asymptotic form of the solution at large $x$ (\ref{eq:asymform}) is unaffected by the singular terms of the potential. It follows that the Wronskians with adjacent subscripts, $W_{k,k+1}$, evaluate to the same expressions as given in (\ref{wneigh}).

The solutions to the Y-system are determined by the asymptotic behavior of the Y-functions in the semi-classical limit $\hbar \to 0$, the latter governed by the ascending paths similarly to the considerations in the previous section. Note that at fixed angular momentum $\ell$, the $x^{-2}$ term in the potential $V(x)$ vanishes in the $\hbar\to 0$ limit. Thus for the purpose of determining the semi-classical limit of Y-functions, it suffices to replace $V(x)$ with the potential
\begin{equation}
	\widetilde V(x) \equiv x^{r+1} + \sum_{a=0}^{r+1} u_{a} x^{r-a} .
\end{equation}
Note that $\widetilde V(x)$ may admit a simple pole at $x=0$. 

\begin{figure}[h!]
	\centering
	\includegraphics[width=0.42\linewidth]{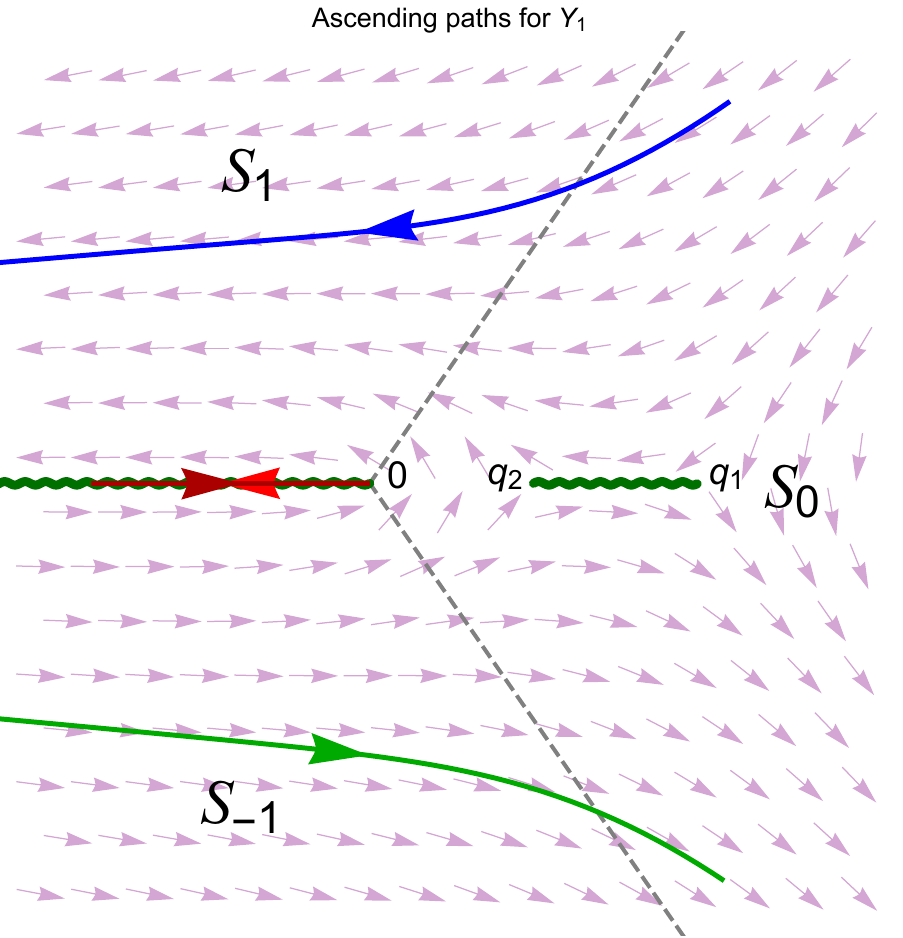}~~~
	\includegraphics[width=0.42\linewidth]{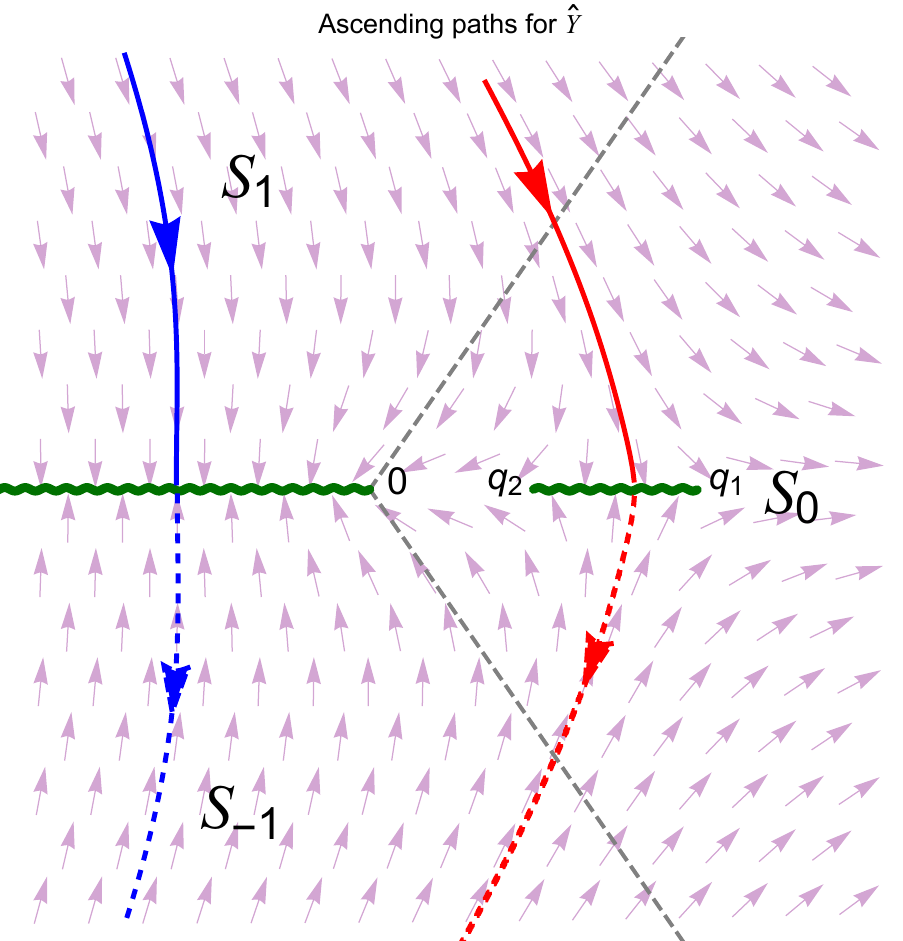}
	\caption{Collections of ascending paths on the complex $x$-plane that determine the semi-classical limit of $Y_1$ (left) and $\hat Y$ (right) respectively, in the $r=0$ case.}
	\label{fig:ascsing}
\end{figure}

We begin by considering the case where the roots of $x\widetilde V(x)=0$, denoted $q_1,\cdots, q_{r+2}$, are positive and distinct, and ordered according to 
\ie
\label{eq:order_qs_sing}
	q_1>q_2>\cdots > q_{r+2} > 0
\fe
The WKB approximation for both the $Y_s$'s and $\hat Y$ are determined by integrating $\sqrt{\widetilde V(x)}$ along suitable collections of ascending paths, similarly to the analysis in the previous section. This is illustrated in Figure \ref{fig:ascsing} in the $r=0$ case.

Equations (\ref{asympeven}) and (\ref{asympodd}), with the replacement $V \to \widetilde V$, are still valid here. In addition, we have
\ie\label{asymphat}
\hat Y(u_a, \hbar) \approx e^{ - {\hat m \over \hbar} },~~~~ \hat m \equiv  i \oint_{\hat \gamma} \sqrt{\widetilde V(x)} dx,
\fe
where $\hat \gamma$ encloses the pole at $x=0$ and $q_{r+2}$. 

From this point on, the derivation of the TBA equations from the Y-system is in closely parallel with the case of polynomial potential, with an important new subtlety. It turns out that the Y-functions (more precisely, their logarithms), depending on the value of $\ell$, may not be analytic in the strip $-{\pi\over 2}<{\rm Im}\theta<{\pi\over 2}$, as we observe numerically in section \ref{sec:NotwallCross}. If we naively make the same analyticity assumption as in the case of polynomial potential, we would derive the TBA equations
\begin{eqnarray}\label{naivetbae}
	\epsilon_s( \theta) &=& m_s e^\theta - K\star L_{s+1}(\theta) - K\star L_{s-1}(\theta) ,\  s=1,\dots,r ,\\
	\epsilon_{r+1}( \theta) &=& m_{r+1} e^\theta - K\star L_{r}(\theta) - K\star \hat L(\theta) ,\\
	\hat \epsilon( \theta) &=& \hat m e^\theta - K\star L_{r+1}(\theta) .
\end{eqnarray}
Here $\epsilon_s$ and $L_s$ are defined as in (\ref{yepsrel}), whereas $\hat\epsilon$ and $\hat L$ are defined by 
\ie
\label{yepsrel2}
	\hat Y(u_a, \hbar) \equiv \exp\left[ -\hat \epsilon(u_a, \theta) \right] ,
	~~~~ \hat L(\theta) \equiv \log\left[\left(1-e^{2\pi i \ell}e^{-\hat \epsilon(\theta)}\right)\left(1-e^{-2\pi i \ell}e^{-\hat \epsilon(\theta)}\right)\right].
\fe
Note that these equations only depend on $\ell$ modulo an integer. Together with the fact that the Y-functions are independent of $\ell$ in the semi-classical limits, this is in clear contradiction with the physical expectation on the spectrum which depends on $\ell$ not just modulo an integer. Thus the analyticity assumption made in the derivation of (\ref{naivetbae}) must fail for general $\ell$. 

Indeed, following a detailed analysis of the $r=0$ case in section \ref{sec:NotwallCross}, we will conjecture that the ``naive" analyticity assumption holds for $-1<\ell<0$, while the extension of the TBA equations to other values of $\ell$ can be derived by a careful consideration of analytic continuation, during which additional terms arise whenever a singularity of $L_{s}$ or $\hat L$ crosses the integration contour.\footnote{A similar phenomenon was observed in \cite{Dorey:1996re} in the context of the spectrum of integrable field theories on a cylinder.} In contrast to the usual wall-crossing phenomenon discussed in the previous section and Appendix \ref{sec:wallCrossing}, where singularities of the kernel $K$ cross the integration contour, here relevant singularities are those of the $L$ functions. In particular, while the usual wall-crossing can be detected at the level of the classical mass parameters (see Appendix \ref{sec:wallCrossing}), here we do not know an a priori way of determining the crossing of singularities; rather, we do so by inspecting the numerical solution of the TBA equations along the path of analytic continuation in $\ell$.

\section{Exact Quantization Conditions}
\label{sec:EQC}

The TBA equations formulated in the previous section determine the quantum periods, or the Y-functions, in terms of the parameters of the Schr\"odinger system including the energy. The spectrum is then determined through the exact quantization condition (EQC) as suitable algebraic conditions on the Y-functions, which we shall formulate in this section.

In this approach, one naturally arrives at the so-called Voros spectrum, which expresses the possible values of $\hbar$ at a given energy. 
The energy spectrum as functions of $\hbar$ can in turn be obtained by inverting the Voros spectrum. We will formulate the EQC in the minimal chamber. Upon wall-crossing, the EQC remains valid; however, practicality requires re-expressing the EQC in terms of the new $Y$-functions appearing in the wall-crossed TBA equations (see Appendix \ref{sec:wallCrossing} and comments at the end of section \ref{sec:Exact}).

\subsection{Polynomial potential}
\label{sec:eqcnonsing}

For a polynomial potential $V(x) = x^{r+1}+\cdots$ with odd $r$, the spectrum of interest is that of bound (i.e. normalizable) states, with real positive $\hbar$. The quantization condition amounts to the normalizability of the wave function $\psi(x)$ (defined on the real $x$-axis). Such $\psi(x)$ must decay both for $x\to +\infty$, or in the wedge ${\cal S}_0^\gg$, and for $x\to -\infty$, or in the wedge ${\cal S}_{r+3\over 2}^\gg$. In other words, $\psi(x)\propto y_0(x)\propto y_{r+3\over 2}(x)$. Thus the EQC is equivalent to
\ie\label{eqcrodd}
	Q \equiv  W_{0,{r+3\over 2}}(u_a, \hbar) = 0 .
\fe
In the case of even $r$, on the other hand, bound states do not exist for real positive $\hbar$. The wave function $\psi(x)$ that decays in $x\to +\infty$ is a superposition of incoming and outgoing waves in $x\to -\infty$. If we analytically continue in $\hbar$ by assigning a small positive phase, the incoming wave is turned into a decaying wave function in the $x\to -\infty$ limit, while the outgoing wave now growing at $x\to -\infty$. When the outgoing wave is absent, the wave function is normalizable and defines a resonance. The EQC for resonances is
\ie\label{eqcreven}
	Q \equiv W_{0, {r+2\over 2}}(u_a, \hbar) 
	 = 0 .
\fe

We would like to express (\ref{eqcrodd}) or (\ref{eqcreven}) as a constraint on the Y-functions, using the Plucker relations (\ref{pluckerr}) as well as the known Wronskians between decaying solutions in neighboring wedges (\ref{wneigh}). 
A simple derivation is given below for the expression of the EQC in terms of the Y-functions.

\begin{figure}[h!]
	\centering
	\includegraphics[width=.8\linewidth]{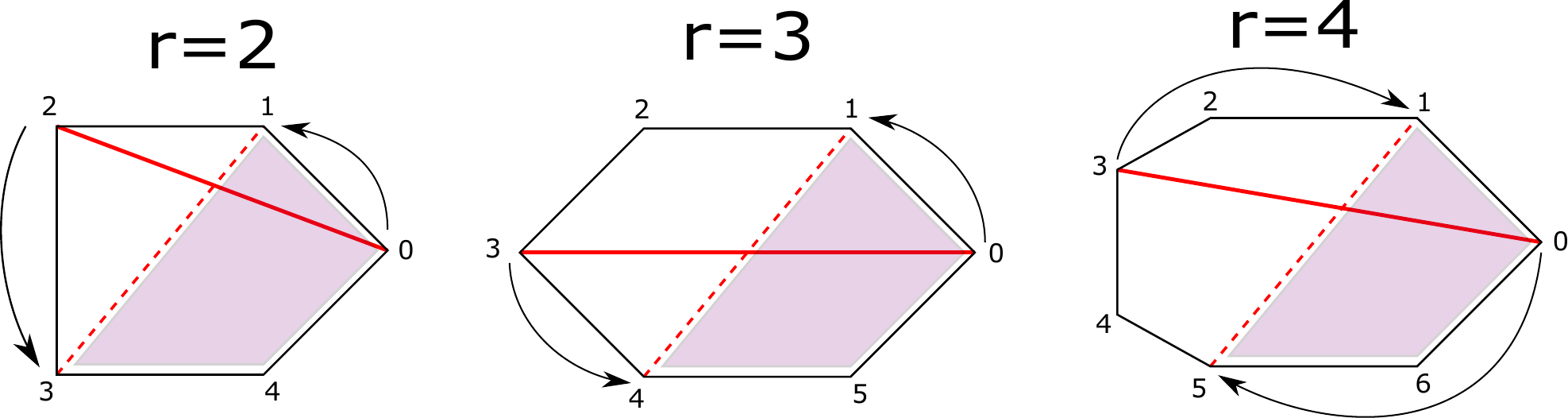}
	\caption{Vertices represent the solutions $y_k$, edges represent the Wronskian of any two solutions. The shaded region represents $Y_1(e^{-i\pi/2}\hbar)$, in the sense that it is a multiplication of the Wronskians on its edges or their inverse. We use the behavior of the Wronkians under a shift of $\hbar$ (\ref{eq:wrelation}), and the triviality of Wronkians between adjacent solutions (\ref{wneigh}) to re-express the EQC in terms of the $Y$-functions.}
	\label{fig:EQC023}
\end{figure}

\paragraph{$\mathbf{r\equiv0,2,3~{\rm \textbf{mod}}~4}$} We begin by using the relation
\begin{equation}
\label{eq:wrelation}
	W_{0,m}(\hbar) = W_{k,m+k}\left(e^{-\pi i k}\hbar\right)
\end{equation}
which follows from (\ref{wkshfone}) (the RHS is defined via the analytic continuation specified by the phase rotation of $\hbar$), and rewrite the EQC as (see figure \ref{fig:EQC023})
\ie\label{qqinit}
	Q = \left\{
	\begin{array}{ll}
	W_{-\frac{r+4}{4},\frac{r}{4}}\left( e^{i\pi {r+4\over 4}} \hbar\right), & \quad r\equiv 0~{\rm mod}~4 \\
	W_{\frac{r+2}{4},\frac{3(r+2)}{4}}\left( e^{-i\pi {r+2\over 4}} \hbar\right), & \quad r\equiv 2~{\rm mod}~4 \\
	W_{\frac{r+1}{4},\frac{3r+7}{4}}\left( e^{-i\pi {r+1\over 4}} \hbar\right), & \quad r\equiv 3~{\rm mod}~4
	\end{array}
	\right. 
\fe
Note that the indices of $W$ are defined mod $r+3$, up to a sign,
\begin{equation}
	W_{i,j+r+3}(\hbar) = -W_{i,j}(\hbar) = W_{j,i}(\hbar)\ .
\end{equation}
Using (\ref{wneigh}) and the definition of the Y-functions as cross ratios of Wronskians, we have
\ie
\label{yfuncsexplicit1}
& Y_{2j+1}(u_a, e^{i\pi \left(k-\frac{1}{2}\right)}\hbar) = 	\left\{
\begin{array}{ll}
	W_{-j-1,j}W_{-j-2,j+1}(u_a, e^{i\pi k}\hbar),  & \quad r ~\text{even} \\
	\omega^{-2(-)^{j+k} B_{\frac{r+3}{2}}}W_{-j-1,j}W_{-j-2,j+1}(u_a, e^{i\pi k}\hbar), & \quad r ~\text{odd}
\end{array}
\right. 
\\
&
Y_{2j}(u_a, e^{i\pi k}\hbar) = 
W_{-j,j}W_{-j-1,j+1}(u_a, e^{i\pi k}\hbar).
\fe
It follows that a Wronskian with an odd difference between its subscripts can be expressed in terms of the Y-functions,
\begin{equation}
\label{eq:odd_putinYs}
	W_{-l-1+k,l+k}(\hbar) =\omega^{(-)^{k} B_{\frac{r+3}{2}}} \prod_{j=0}^{l-1}\left[\omega^{2(-)^{j+k+1} B_{\frac{r+3}{2}}}Y_{2j+1}\left(\theta-i\pi \left(k-1/2\right)\right)\right]^{(-1)^{j+l-1}}\ .
\end{equation}
Here we have adopted a notation for the Y-functions in which the $u_a$-dependence is omitted, and the argument is the spectral parameter $\theta$ rather than $\hbar$ (related by $\hbar = e^{-\theta}$).
It is then possible to rewrite (\ref{qqinit}) as a ratio of products of Y-functions, 
\ie
\label{eq:Q023}
	Q = \left\{
	\begin{array}{ll}
		\prod_{j=0}^{{r\over 4}-1} \left[ Y_{2j+1}\left(\theta-i\pi\left( \frac{r}{4} +\frac{1}{2}\right) \right) \right]^{(-)^{j+{r\over 4}+1} } , & \quad r\equiv 0~{\rm mod}~4 \\
		\prod_{j=0}^{\lfloor \frac{r}{4} \rfloor} \left[ Y_{2j+1}\left(\theta + i\pi \left(\lfloor \frac{r}{4} \rfloor+\frac{3}{2}\right) \right) \right]^{(-)^{j+\lfloor \frac{r}{4} \rfloor} }, & \quad r\equiv 2~{\rm mod}~4 \\
		\omega^{(-)^{\lfloor \frac{r}{4} \rfloor+1}B_{\frac{r+3}{2}}}\prod_{j=0}^{\lfloor \frac{r}{4} \rfloor} \left[\omega^{2(-)^{j+\lfloor \frac{r}{4} \rfloor +1} B_{\frac{r+3}{2}}} Y_{2j+1}\left(\theta+i\pi \left(\lfloor \frac{r}{4} \rfloor+\frac{3}{2}\right) \right)\right]^{(-)^{j+\lfloor \frac{r}{4} \rfloor}}, & \quad r\equiv 3~{\rm mod}~4
	\end{array}
	\right.
\fe

To make contact with the TBA equations, it is more convenient to express the EQC in terms of the values of the Y-functions on the strip $-{\pi\over 2}\leq {\rm Im} \theta \leq {\pi\over 2}$. Indeed, using the relation 
\ie
\label{eq:Yshift}
&	Y_{s}(\theta+i \pi k) = \left[ Y_{s}(\theta) \right]^{(-)^k} \prod_{\pm,j=0}^{k-{\rm sgn}(k)} \left[1+Y_{s\pm 1}\left(\theta+i\pi j + i\pi {{\rm sgn}(k)\over 2} \right)\right]^{(-)^{j+k-1}},
\fe
which may be derived by repeatedly applying (\ref{ysystem}), we can express the EQC in terms of $Y_s(\theta)$ for real $\theta$ at even $s$, and for $\theta\in \mathbb{R} \pm {i\pi\over 2}$ fo odd $s$.

\begin{figure}[h!]
	\centering
	\includegraphics[width=.8\linewidth]{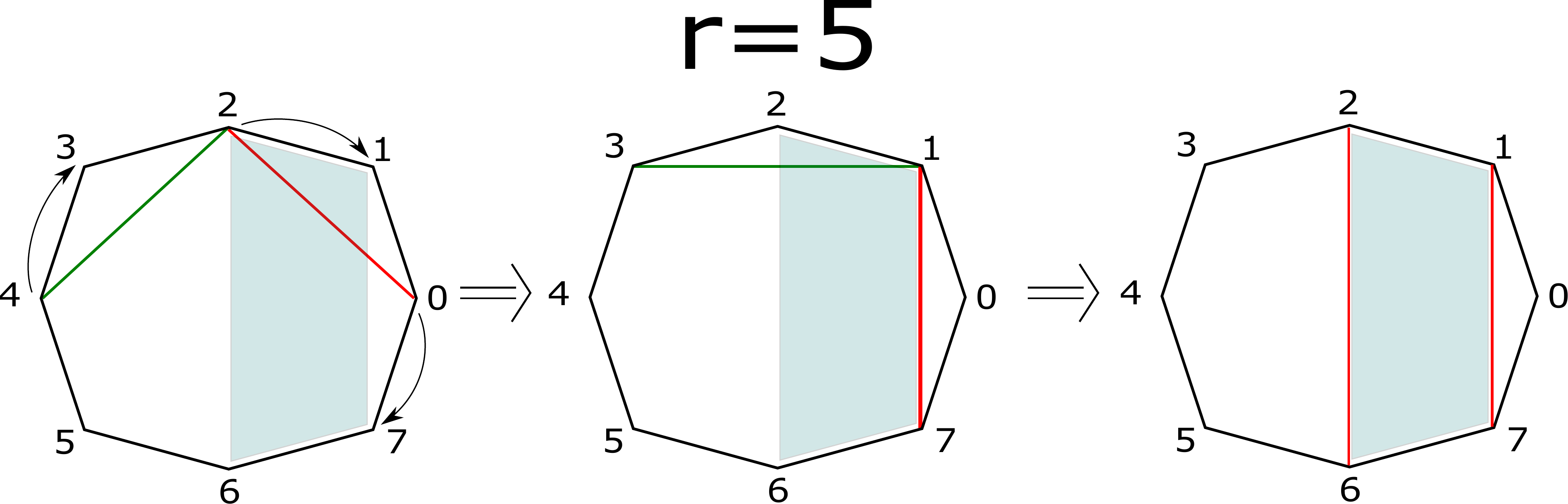}
	\caption{The shaded region here represents $Y_2$. See the caption of figure \ref{fig:EQC023} for other conventions. For $r=5$, the LHS of equation (\ref{eq:EQC1mod4}) has the ratio of the Wronskians corresponding to the edges in red to the ones in gree n. First we shift $\hbar$ to rotate them. Second, we use a plucker relation to exchange $W_{1,3}$ with $W_{2,6}$. Of course, the plucker relation introduces additional factors of Wronskians, but we omit them from the figure since they can be put in terms of $Y$-functions employing the approach used in the ${r\equiv0,2,3~{\rm {mod}}~4}$ cases. On the right we see that we are left with the ratio of Wronskians that can be expressed in terms of $Y_2$}
	\label{fig:EQC1}
\end{figure}

\paragraph{$\mathbf{r\equiv1~{\rm \textbf{mod}}~4}$}
Using the identity 
$
	y_{j} = \frac{W_{j,(r-1)/2}}{W_{(r+1)/2,(r-1)/2}} y_{(r+1)/2} - \frac{W_{j,(r+1)/2}}{W_{(r+1)/2,(r-1)/2}} y_{(r-1)/2} ,
$
the EQC which amounts to $y_0 \propto y_{r+3\over 2}$ can be written as (assuming $r>1$)
\begin{equation}
\label{eq:EQC1mod4}
	\frac{W_{0,\frac{r-1}{2}}}{W_{\frac{r+3}{2},\frac{r-1}{2}}} = \frac{W_{0,\frac{r+1}{2}}}{W_{\frac{r+3}{2},\frac{r+1}{2}}} .
\end{equation}
The RHS of (\ref{eq:EQC1mod4}) is a Wronskian of the type (\ref{eq:odd_putinYs}), and hence can be expressed in terms of Y-functions as
\ie
\label{eq:case1RHS}
	\frac{W_{0,\frac{r+1}{2}}}{W_{\frac{r+3}{2},\frac{r+1}{2}}} &= -W_{-\frac{r+3}{4},\frac{r-1}{4}}\left(e^{\pi i \frac{r+3}{4}} \hbar\right)\omega^{B_{\frac{r+3}{2}}}
\\
&=\omega^{\left(1+(-1)^{ \frac{r-1}{4} }\right)B_{\frac{r+3}{2}}}\prod_{j=0}^{\frac{r-5}{4}} \left[\omega^{2(-)^{j+ \frac{r-1}{4}  } B_{\frac{r+3}{2}}} Y_{2j+1}\left(\theta-i\pi \left( \frac{r-1}{4} +\frac{1}{2}\right) \right)\right]^{(-)^{j+ \frac{r-5}{4}}}\ .
\fe
The LHS of  (\ref{eq:EQC1mod4}) can be expressed using Plucker relation as
\ie
\label{eq:case1LHS}
	\frac{W_{0,\frac{r-1}{2}}}{W_{\frac{r+3}{2},\frac{r-1}{2}}}(\hbar) &= -\frac{W_{-\frac{r-1}{4},\frac{r-1}{4}}}{W_{\frac{r-1}{4},\frac{r+7}{4}}}\left(e^{i\pi \frac{r-1}{4}}\hbar \right) 
\\
&= -\frac{W_{-\frac{r-1}{4},\frac{r-1}{4}}W_{-\frac{r+3}{4},\frac{r+3}{4}}}{W_{\frac{r-1}{4},\frac{r+3}{4}}W_{-\frac{r+3}{4},\frac{r+7}{4}} - W_{\frac{r+7}{4},\frac{r+3}{4}}W_{-\frac{r+3}{4},\frac{r-1}{4}}}\left(e^{i\pi \frac{r-1}{4}}\hbar \right) .
\fe
In the last line, the denominator can again be expressed in terms of Y-functions using (\ref{eq:odd_putinYs}), whereas the numerator is equal to $Y_{\frac{r-1}{2}}\left(e^{i\pi \frac{r-1}{4}}\hbar \right)$ by (\ref{yfuncsexplicit1}) (see also figure \ref{fig:EQC1}).
Therefore, (\ref{eq:EQC1mod4}) can be expressed solely in terms of the Y-functions, as desired. For numerical application, we can further apply the relation \ref{eq:Yshift} to shift the arguments of the Y-functions appearing in the EQC to the strip $|Im[\theta]|\le \pi/2$.

\subsection{Potential with a regular singularity}
\label{sec:eqcsing}

For the Schr\"odinger problem with a potential of the form $V(x) = {\hbar^2 \ell(\ell+1)\over x^2}+{u_{r+2} \over x}+\cdots+x^{r+1}$, the admissible wave functions should vanish as $x\to +\infty$, and have one of the two possible behaviors (\ref{ynearorign}) near $x=0$. To write the EQC amounts to recasting these boundary conditions on the wave functions in terms of the quantum periods or the Y-functions.

In addition to the Plucker relations (\ref{pluckerr}), we will make use of the identities
\ie
	W_{i+r+3,j+r+3}(\hbar) = W_{ij}(\hbar) ,~~~~ 
	W_{j+r+3,\pm}(\hbar) = -e^{\pm 2\pi i\ell} W_{j\pm}(\hbar),
\fe
where the subscript $\pm$ of the Wronskian refers to the basis of solutions $y_\pm$ (\ref{ynearorign}).
If $2\ell$ is not an integer, $y_+$ and $y_-$ are distinguished by their monodromies around $x=0$. We can expand the decaying solution in the wedge ${\cal S}_0$ as
\ie
	y_0(x) = \frac{W_{0+}}{W_{-+}} y_-(x) + \frac{W_{0-}}{W_{-+}} y_+(x) ,
\fe
where we omit argument dependence on $\hbar$ and $u_a$.
After undergoing the monodromy, we have
\ie\label{eq:r+3to0}
	y_{r+3}(x) = - y_0(e^{-2\pi i}x) = - \frac{1}{W_{-+}}\left(e^{-2 \pi  i \ell} W_{0+} y_- + e^{2 \pi  i \ell} W_{0-} y_+\right) .
\fe
The EQC amounts to either $W_{0+}=0$ or $W_{0-}=0$, depending on the choice of boundary condition at $x=0$. This is equivalent to
\ie
	y_{r+3}(x,u_a,\hbar) = - e^{\pm 2 \pi i \ell} y_0(x,u_a,\hbar).
\fe
Using the basis of any pair of decaying solutions in nearby wedges, $y_{k-1},y_{k}$, we can express the EQC in terms of the Wronskians as
\ie \label{eq:EQCsingMomdromy}
&	W_{k,r+3}(\hbar) = - e^{\pm 2 \pi i \ell} W_{k,0}(\hbar) , ~~~~ W_{k-1,r+3}(\hbar) = - e^{\pm 2 \pi i \ell} W_{k-1,0}(\hbar).
\fe
Next, we shall convert these relations to those of Y-functions.

\paragraph{even $r$}
We choose $k=r+3$ and obtain from the first condition of (\ref{eq:EQCsingMomdromy}),
\ie
\label{eq:sing_even_cond_1}
	0 = W_{r+3,0}(\hbar) = W_{\frac{r+2}{2},-\frac{r+4}{2}}(e^{i\pi\frac{r+4}{2}}\hbar), 
\fe
which can be expressed in terms of the Y-functions with odd indices using (\ref{eq:odd_putinYs}) (still valid here). The second condition of (\ref{eq:EQCsingMomdromy}) can be put in the form
\ie\label{seceven}
	{W_{r+2,r+3}}(\hbar) = e^{\pm 2 \pi i \ell} \hat Y(e^{i\pi(r+2)/2}\hbar)W_{-1,0}(\hbar) ,
\fe
using the definition of $\hat Y$ (\ref{eq:Yhat}). The Wronskians with adjacent subscripts are then determined through (\ref{wneigh}).

\paragraph{odd $r$}
We choose $k=r+2$. The first condition of (\ref{eq:EQCsingMomdromy}) then leads to (\ref{seceven}). The second condition of (\ref{eq:EQCsingMomdromy}) is
\ie\label{secodd}
W_{r+1,r+3} &= - e^{\pm 2 \pi i \ell} W_{r+1,0} .
\fe
Using Plucker relations, we can write
\ie
\frac{W_{r+1,r+3}}{W_{r+1,0}}(\hbar) &= -\frac{W_{\frac{r+1}{2},\frac{r+5}{2}}}{W_{-\frac{r+1}{2},\frac{r+1}{2}}}(e^{i \pi (r+1)/2}\hbar) 
\\
&=  -\frac{W_{\frac{r+1}{2},\frac{r+3}{2}}W_{-\frac{r+3}{2},\frac{r+5}{2}}-W_{\frac{r+5}{2},\frac{r+3}{2}}W_{-\frac{r+3}{2},\frac{r+1}{2}}}{W_{-\frac{r+3}{2},\frac{r+3}{2}} W_{-\frac{r+1}{2},\frac{r+1}{2}}}(e^{i \pi (r+1)/2}\hbar)   ,
\fe
and thus put (\ref{secodd}) in the form
\begin{equation}
	-\frac{W_{\frac{r+1}{2},\frac{r+3}{2}}W_{-\frac{r+3}{2},\frac{r+5}{2}}-W_{\frac{r+5}{2},\frac{r+3}{2}}W_{-\frac{r+3}{2},\frac{r+1}{2}}}{Y_{r+1}}(e^{i \pi (r+1)/2}\hbar) = \omega^{(r+3)/2} e^{\pm 2 \pi i \ell}\ .
\end{equation}
Now the numerator involves only Wronskians with odd differences in their subscripts, which can be expressed solely in terms of the Y-functions using (\ref{eq:odd_putinYs}).

Finally, to make contact with solutions to the TBA equation, we can use (\ref{eq:Yshift}) for $Y_{0<s<r+1}$, together with
\ie
\label{eq:Yshiftsing}
&	Y_{r+1}(\theta+i \pi k) = \left[ Y_{r+1}(\theta) \right]^{(-)^k} \prod_{j=0}^{k-{\rm sgn}(k)} \left[1+Y_{r}\left(\theta+i\pi j + i\pi {{\rm sgn}(k)\over 2} \right)\right]^{(-)^{j+k-1}}\nonumber \\
&\times \prod_{j=0}^{k-{\rm sgn}(k)} \left[\left(1-e^{2\pi i \ell}\hat Y\left(\theta+i\pi j + i\pi {{\rm sgn}(k)\over 2} \right)\right)\left(1-e^{-2\pi i \ell}\hat Y\left(\theta+i\pi j + i\pi {{\rm sgn}(k)\over 2} \right)\right)\right]^{(-)^{j+k-1}} \ ,\\
&	\hat Y(\theta+i \pi k) = \left[ \hat Y(\theta) \right]^{(-)^k} \prod_{j=0}^{k-{\rm sgn}(k)} \left[1+Y_{r+1}\left(\theta+i\pi j + i\pi \frac{{\rm sgn}(k)}{2}\right)\right]^{(-)^{j+k-1}}  ,
\fe
to shift the arguments of the Y-functions appearing in the EQC to the strip $|Im[\theta]|\le \pi/2$.

\section{Examples with Polynomial Potential}
\label{sec:EXpol}
In this section we review some examples of Schr\"odinger systems with a polynomial potential, for which we write down the explicit EQC following the general prescription of section \ref{sec:EQC}. 

As explained below (\ref{eq:Yshift}), the EQC can be expressed in terms of $Y_s(\theta)$ with real $\theta$ for even $s$, and $\theta\in \mathbb{R}\pm {i\pi\over 2}$ for odd $s$. Here we are using the notation for $Y$ adopted in (\ref{eq:odd_putinYs}), in which the argument is $\theta\equiv -\log(\hbar)$. For brevity, we will omit the arguments of the $Y$'s when there is no room for confusion, denoting $Y_s \equiv Y_s(\theta)$ and $Y_s^{\pm} \equiv Y_s^{\pm}(\theta) \equiv Y_s(\theta \pm i\pi/2)$. 
In some cases, we will follow a convention common in the literature and express the EQC using $Y_s^{med}(\theta) \equiv \sqrt{Y_s^+(\theta)/Y_s^-(\theta)}$.

\subsection{Harmonic oscillator ($r=1$)}

The derivation given in section \ref{sec:eqcnonsing} is not applicable to the harmonic oscillator (i.e. $r=1$), as noted around (\ref{eq:EQC1mod4}). This can be amended by considering a slightly modified ``necessary EQC" (nEQC) $W_{0,2}W_{1,3}(\hbar)=0$, and in the end we can select the physical spectrum out of the candidate spectrum allowed by the nEQC. 
The Y-system in this case is rather trivially written as
\begin{equation}
	Y_1(\theta+i\pi/2) = e^{i\pi u_1 e^{\theta}} ,
\end{equation}
and is not needed. Using a Plucker relation and the Wronskian between decaying solutions of adjacent wedges, we can express the nEQC as
\ie\label{wwneqc}
	W_{0,2}W_{1,3} = 2 \cos \left(\frac{\pi  u_1}{2 \hbar }\right)=0,
\fe
where $u_1$ is minus the energy. The solution to (\ref{wwneqc}),
\begin{equation}
\label{eq:quantHO}
	{u_1} \in (1 + 2\mathbb{Z})\hbar ,
\end{equation}
is the Bohr-Sommerfeld quantization condition without restricting to positive energies. If we further make the mild assumption of positive energy spectrum when $\hbar > 0$, the Schr\"odinger problem on the imaginary $x$-axis then has negative energy spectrum, i.e. the solution to $W_{1,3}=0$ has $u_1>0$, leaving the entirety of $u_1<0$ solutions of (\ref{eq:quantHO}) to be the actual physical spectrum.

\subsection{Cubic potential ($r=2$)}

As an example of the EQC for general polynomial potentials derived in section \ref{sec:eqcnonsing}, let us inspect the $r=2$ case, namely the EQC for a resonance in a cubic potential.
The latter can be expressed in terms of the Y-functions following (\ref{eqcreven})-(\ref{qqinit}) as
\ie
0= W_{0,2}(\hbar) = W_{1,3}(e^{-i\pi}\hbar) = Y_{1}\left(\theta-{3 i \pi\over 2}\right) .
\fe
Note that we are using the notation for $Y$ adopted in (\ref{eq:odd_putinYs}), in which the argument is $\theta\equiv -\log(\hbar)$.
Applying (\ref{eq:Yshift}) repeatedly, we can write the RHS as
\ie
Y_{1}\left(\theta-{3 i \pi\over 2}\right) &= \frac{1+Y_2(\theta-i\pi)}{Y_1^-(\theta)} = \frac{1+Y_1^-(\theta) + Y_2(\theta)}{Y_1^-(\theta) Y_2(\theta)} ,
\fe
where $Y_1^-$ is defined at the beginning of this section. We can use the Y-system equations (\ref{ysystem}), or equivalently (\ref{eq:Yshift}), to relate
\ie
\label{eq:Ysystemplusminus}
& Y_1^+ Y_1^- = 1+Y_2,
\fe
thereby rewriting the EQC in the following two equivalent forms,
\begin{equation}\label{eq:cubicEQCreso}
0= W_{0,2}(\hbar) = \frac{1+Y_1^+(\theta) + Y_2(\theta)}{Y_1^+(\theta) Y_2(\theta)} = \frac{Y_1^{med}(\theta)+\sqrt{1+Y_2(\theta)}}{Y_1^{med}(\theta) Y_2(\theta)} .
\end{equation}
For real parameters of the potential $V(x)$ (including the energy) in the minimal chamber and real $\theta$ (i.e. real positive $\hbar$), $Y_1^{med}(\theta)$ is a phase and $Y_2(\theta)>0$. As expected, the EQC (for resonance) cannot be satisfied, unless the energy is taken to be complex. Numerical analysis of this EQC have been performed in \cite{Emery:2020qqu,Ito:2018eon}.

\subsection{Quartic potential ($r=3$)}
\label{sec:quadcase} 

For the bound state problem in a quartic potential, the EQC is given by (\ref{eq:Q023}) in the case $r=3$, namely
\begin{equation}
0=Q=\omega^{-3B_{3}} Y_{1}\left(\theta+\frac{3}{2}i\pi \right).
\end{equation}
Next, we can use (\ref{eq:Yshift}) to re-express this relation as
\ie \label{eq:quartEQC}
\frac{Y_1^+(\theta) Y_3^+(\theta)+Y_1^+(\theta)+Y_2(\theta)+Y_3^+(\theta)+1}{Y_1^+(\theta) Y_2(\theta)} e^{i \pi  B_3} = 0,
\fe
so that for real $\theta$, the EQC only involves Y-functions whose arguments lie in the strip $-\pi/2<{\rm Im} (\theta)\le \pi/2$.

Alternatively, this relation can be recast in terms of $Y_2(\theta)$ and $Y_s^{\rm med}(\theta)$ as
\begin{equation}\label{eq:quartEQCmed}
\frac{{Y_1^{med}}(\theta)+{Y_3^{med}(\theta)}+(1+Y_1^{med}(\theta) Y_3^{med}(\theta)){\sqrt{Y_2(\theta)+1}}}{Y_1^{med}(\theta)Y_2(\theta)}e^{i \pi  B_3} = 0\ .
\end{equation}
This EQC has been analyzed numerically in the case of an even quartic potential in \cite{Emery:2020qqu,Ito:2018eon}.

\begin{figure}[h!]
	\centering
	\includegraphics[width=0.5\linewidth]{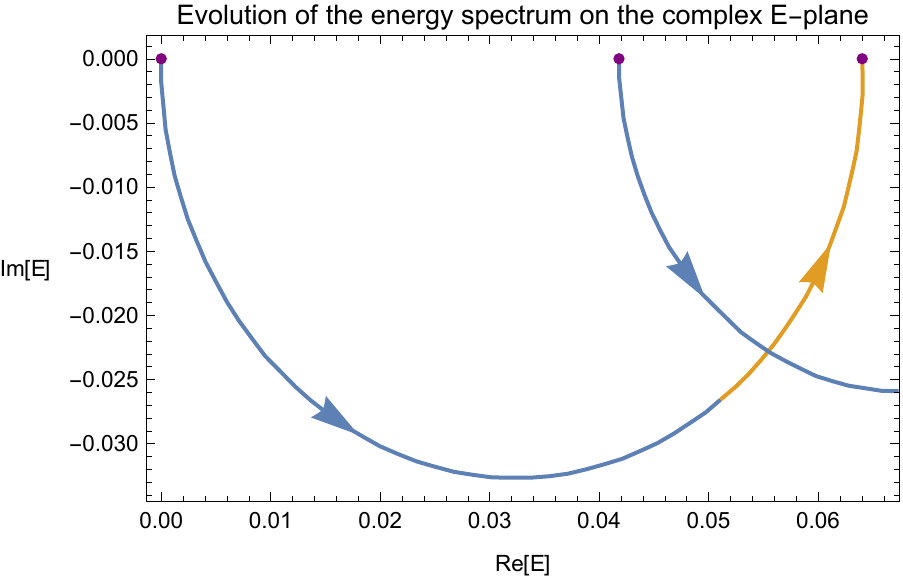}
	\caption{The evolution of the energy spectrum of the Schr\"odinger problem (\ref{eq:schr}) with the quartic potential $V(x)=x^4-\frac{x^2}{2}+\frac{91}{1200}+u_2 x$ under analytic continuation of $u_2$. Here we take $u_2=0.07e^{i\phi}$ and vary $\phi$ from $0$ to $\pi$. The paths of the lowest two energy eigenvalues, computed from the TBA equations (\ref{tbaeqns}) and the EQC (\ref{eq:quartEQC}), are shown. The color change from blue to yellow represents the wall crossing from the vacuum chamber to the minimal chamber.
}
	\label{fig:hrotation}
\end{figure}

As an application, let us consider an asymmetric double well quartic potential $V(x)=x^4 + u_1 x^2 + u_2 x + u_3$, with $u_1<0$ and sufficiently small real $u_2$. For sufficiently small $\hbar$, the ground state corresponds to a wave function localized around the global minimum of $V(x)$. One may ask what happens to the ground state when we analytically continue $u_2$ to minus itself, under which the global minimum of $V(x)$ is exchanged with a higher local minimum. The answer, at the level of spectrum, is that the ground state is analytically continued into a distinguished excited energy eigenstate, that is the quantum analog of the classical higher local minimum.
This is demonstrated using the numerical solution to the EQC in Figure \ref{fig:hrotation}. 

In this numerical implementation, the TBA and the EQC a priori gives the Voros spectrum, i.e. $\hbar$ as a function of the energy. We then invert this relation numerically by sampling over a grid of energy values. Furthermore, the Schr\"odinger problem with energies below the higher local minimum of $V(x)$ does not satisfy (\ref{minorder}) and lies outside of the minimal chamber. In this case an additional chamber is required, which we refer to as the vacuum chamber. The TBA+EQC in the vacuum chamber can be found by a straightforward analytic continuation of the equations from the minimal chamber, as demonstrated in Appendix \ref{sec:wallCrossing}.

\subsection{Sextic potential ($r=5$)}

For the bound state problem in a sextic polynomial potential ($r=5$), we can start from the EQC for general ${r\equiv1~{\rm {mod}}~4}$, given by setting (\ref{eq:case1RHS}) to be equal to (\ref{eq:case1LHS}),
\begin{equation}
\label{eq:EQCr_5_raw}
	-\omega^{-4 B_{4}} \frac{Y_1\left(\theta - \frac{3}{2}i\pi\right)Y_2\left(\theta - i\pi\right)}{Y_1\left(\theta - \frac{3}{2}i\pi\right) Y_1\left(\theta - \frac{1}{2}i\pi\right)+Y_3\left(\theta - \frac{3}{2}i\pi\right)} = -\omega^{4B_4} Y_1\left(\theta - \frac{3}{2}i\pi\right) .
\end{equation}
Using (\ref{eq:Yshift}) to shift the arguments of the Y-functions, we can rewrite this relation in terms of $Y_s(\theta)$ an $Y_s^+(\theta)$ as
\begin{equation}
\label{eq:EQCr_5}
	\frac{\omega ^{-4 B_4} \left(\frac{\omega ^{8 B_4} (Y_1^++Y_2+1) (Y_4+Y_5^++1)}{Y_1^+ Y_5^+ (Y_5^++1)}+Y_1^++1\right)}{Y_2} = 0 .
\end{equation}
Here we have omitted the argument $\theta$ in all of the functions involved.\footnote{We have eliminated $Y_3^+$ using the algebraic identity
$
	Y^+_3 = \frac{Y_1^+ W_{0,1}W_{-2,-1} Y_5^+ W_{2,3}W_{-4,-3} }{W_{1,2}W_{-3,-2}W_{-1,0}W_{3,4}} = \omega^{-8B_4} Y^+_1 Y^+_5
$
which follows from (\ref{yfuncs}) and \ref{wneigh}.} To the best of our knowledge, only a very specific symmetric case of this EQC appears in the literature \cite{Emery:2020qqu}, where it is also numerically tested.\footnote{This was a typo in equation (4.32) of \cite{Emery:2020qqu}; as confirmed by its author, the numerical tests in \cite{Emery:2020qqu} were performed with the correct EQC which agrees with (\ref{eq:EQCr_5}).}

\section{Examples with Regular Singularity}
\label{sec:EXsing}

Now we turn to Schr\"odinger system with a potential $V(x)$ that has a regular singularity, namely $x^{-2}$ and $x^{-1}$ terms, in addition to a polynomial of degree $r+1$ in $x$ (\ref{eq:schrSing}). After warming up with the hydrogen atom in the EQC approach, we will focus on the first nontrivial example following the prescription of section \ref{sec:eqcsing}, namely when the polynomial part of the potential is linear (``the non-relativistic meson"). Importantly, we investigate the analytic continuation of the TBA (\ref{naivetbae}) to $\ell>0$, as well as the analytic structure of the Voros spectrum as a function of $\ell$ close to real interval $\ell\in(-1,0)$.

\subsection{The hydrogen atom ($r=-1$)}

The familiar non-relativistic hydrogen atom in the angular momentum $\ell$ sector is equivalent to the Schr\"odinger system with regular singularity in the case $r=-1$, namely
\ie \label{eq:schrSing_r=-1}
	\left(-\hbar^2 \partial_x^2 + 1 + u_{0} x^{-1} + \frac{\hbar^2 \ell(\ell+1)}{x^2}\right) \psi(x) = 0 .
\fe
Following the convention of (\ref{eq:schrSing}), here we take the energy to be $-1$ and the coefficient of the electric potential to be $u_0$. The $Y$-system is ``too trivial" as in the case of the harmonic oscillator. Nonetheless, we can derive a ``necessary EQC" from the monodromy around the origin of the complex $x$-plane,
\begin{equation}
\label{eq:nEQCHydro}
	W_{1,2}(\hbar)-\omega^{(r+3)/2}e^{2\pi i \ell}W_{1,0}(\hbar) = -2 i e^{i \pi  \ell} \sin \left(\pi  \ell+\frac{\pi  u_0}{2 \hbar }\right) = 0 ,
\end{equation}
whose solutions are
\ie\label{eq:allsolHydro}
\hbar = \frac{-u_0}{2(\ell+n)},~~~n\in\mathbb{Z}.
\fe
The condition (\ref{eq:nEQCHydro}) can be rewritten as
\begin{equation}
	W_{0+}(\hbar)W_{1-}(\hbar)=0,
\end{equation}
in the notation of section \ref{sec:eqcsing}. It contains the spectrum of the desired quantization condition $W_{0+}(\hbar)=0$ as well as that of the ``unwanted" condition $W_{1-}(\hbar)=0$.
As in the harmonic oscillator case, one can argue that the solutions to $W_{0+}(\hbar)=0$ in the case $u_0<0$, $\hbar>0$, are simply given by (\ref{eq:allsolHydro}) restricted to positive integer $n$.

\subsection{The non-relativistic meson ($r=0$)}
\label{sec:nonrelMeson}

A non-relativistic model of mesons consists of two particles connected by a string of constant tension, spinning in three dimensional space. In the angular momentum $\ell$ sector, this is described by the Schr\"odinger system (\ref{eq:schrSing}) with $r=0$, namely
\ie \label{eq:schrmeson}
	\left(-\hbar^2 \partial_x^2 + x+u_0 + {u_1\over x} + \frac{\hbar^2 \ell(\ell+1)}{x^2}\right) \psi(x) = 0 .
\fe
In this subsection we analyze the EQC of this system and explore the analytic continuation of its spectrum in $\ell$.

\subsubsection{EQC}

The EQC (\ref{eq:sing_even_cond_1}) and (\ref{seceven}) in the $r=0$ case are
\ie
\label{eq:sing_even_cond_1_r=0}
0 &= W_{2,3}(\hbar)-e^{2 \pi i \ell} \hat Y(\theta-i\pi)W_{-1,0}(\hbar) 
\\
&= \frac{\left(\hat Y-e^{2 i \pi  \ell}\right) \left(1+Y_1^+-e^{2 i \pi  \ell} \hat Y\right)}{\hat Y Y_1^+},
\fe
and
\ie\label{eq:sing_even_cond_2_r=0}
	0 &= e^{2 \pi i \ell} W_{-2,1}(e^{2\pi i}\hbar) = e^{2 \pi i \ell} Y_1\left(\theta-\frac{3}{2}\pi i\right) 
	\\
	&= e^{2 i \pi  \ell} \frac{\left(1+Y_1^+-e^{2 i \pi  \ell} \hat Y\right) \left(1+Y_1^+ -e^{-2 i \pi  \ell}\hat Y\right)}{\hat Y^2 Y_1^+},
\fe
where we have made use of (\ref{yfuncsexplicit1}), (\ref{eq:Yshift}), (\ref{eq:Yshiftsing}). In all of the Y-functions appearing on the RHS of the last equality, the argument $\theta=-\log \hbar$ has been omitted (this notation is explained at the beginning of section \ref{sec:EXpol}). 

Upon analytic continuation to wall-crossed chambers (see the comments in the end of section \ref{sec:ExactPoly} and appendix \ref{sec:wallCrossing}, or a more thorough discussion in \cite{Ito:2018eon}), the denominators appearing on the RHS of (\ref{eq:sing_even_cond_1_r=0}) and (\ref{eq:sing_even_cond_2_r=0}) may diverge, giving rise to extra solutions to the EQC. Nonetheless, when restricted to the minimal chamber, we can take the EQC to be the naive intersection of the two equations,
\begin{equation}
	1+Y_1^+(\theta) -e^{2 i \pi  \ell} \hat Y(\theta) = 0.
\end{equation}

\subsubsection{the TBA for $\ell>0$}
\label{sec:NotwallCross}
As was alluded to at the end of section \ref{sec:exact_sing}, the TBA equation considered in \cite{Ito:2020ueb} and described in section \ref{sec:exact_sing} is observed to hold only for $-1<\ell<0$. In this subsection we extend the prescription to find the correct TBA equations for a potential with a regular singularity and a linear term (defined by (\ref{eq:schrSing}) with $r=0$), in a certain range of $\ell>0$.

We begin with the TBA (\ref{naivetbae}) for $-1<\ell<0$ and $r=0$, of the form
\ie
\label{eq:TBAnaiveell}
	\epsilon_1(\theta) &= m_1e^{\theta} - \int\limits_{\mathbb{R}} \frac{d\theta'}{2\pi} \frac{\log\left((1-e^{2\pi i \ell}\hat Y(\theta'))(1-e^{-2\pi i \ell}\hat Y(\theta'))\right)}{\cosh\left(\theta-\theta'\right)},\\
	\hat \epsilon(\theta) &= \hat m e^{\theta} - \int\limits_{\mathbb{R}} \frac{d\theta'}{2\pi} \frac{\log\left(1+Y_1(\theta')\right)}{\cosh\left(\theta-\theta'\right)}  ,
\fe
where $Y_1(\theta) \equiv e^{-\epsilon_1(\theta)}$, $\hat Y(\theta) \equiv e^{-\hat\epsilon(\theta)}$. These equations are valid in the strip $-{\pi\over 2}<{\rm Im}\theta<{\pi\over 2}$. Now we would like to analytically continue the solutions in $\ell$. In doing so, the analogous TBA equations obeyed by the analytically continued solutions are expected to be corrected from (\ref{eq:TBAnaiveell}) due to singularities of the integrand crossing the $\theta'$-integration contour on the RHS.

\begin{figure}[h!]
	\centering
	\includegraphics[width=0.43\linewidth]{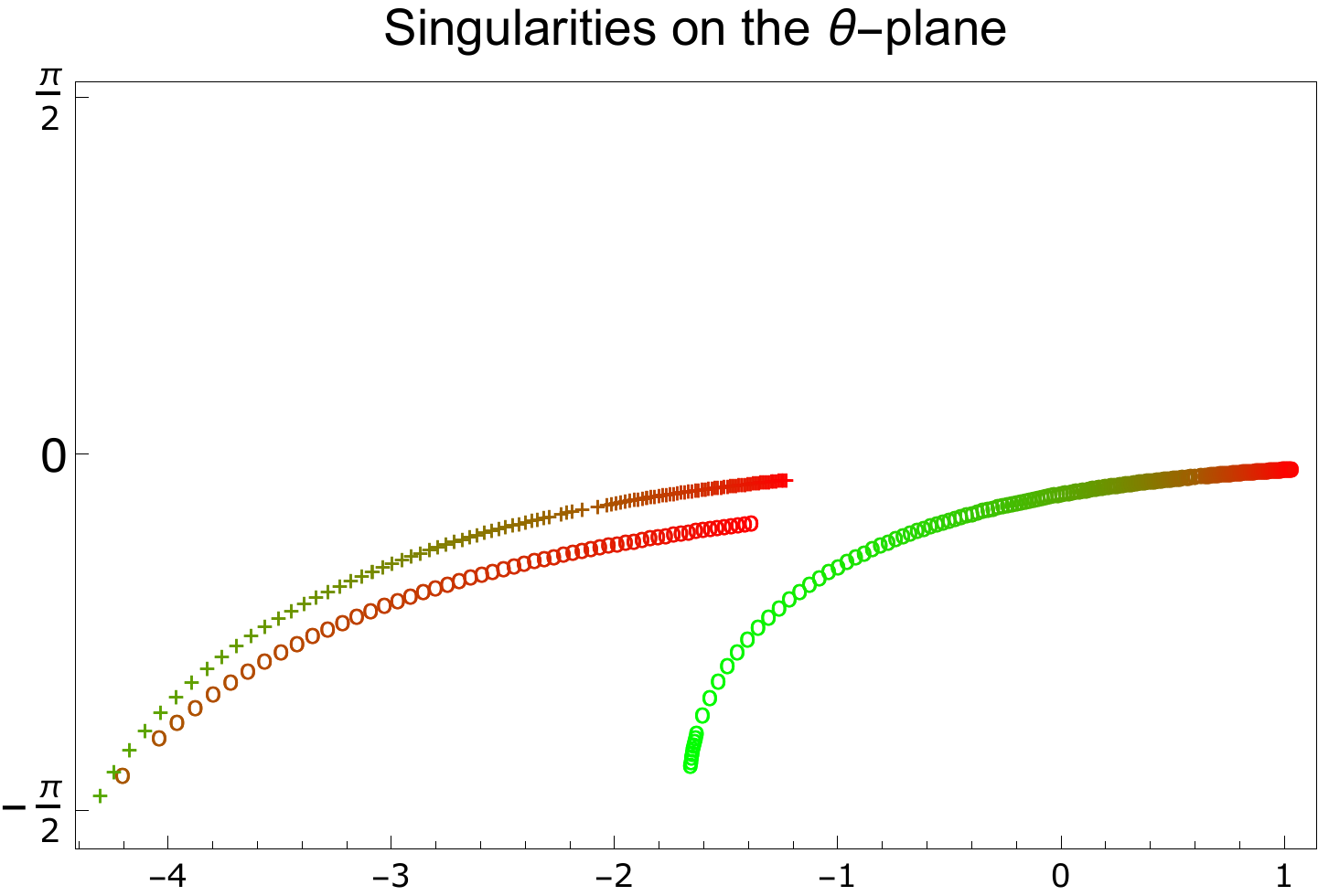}~~~~
	\includegraphics[width=0.48\linewidth]{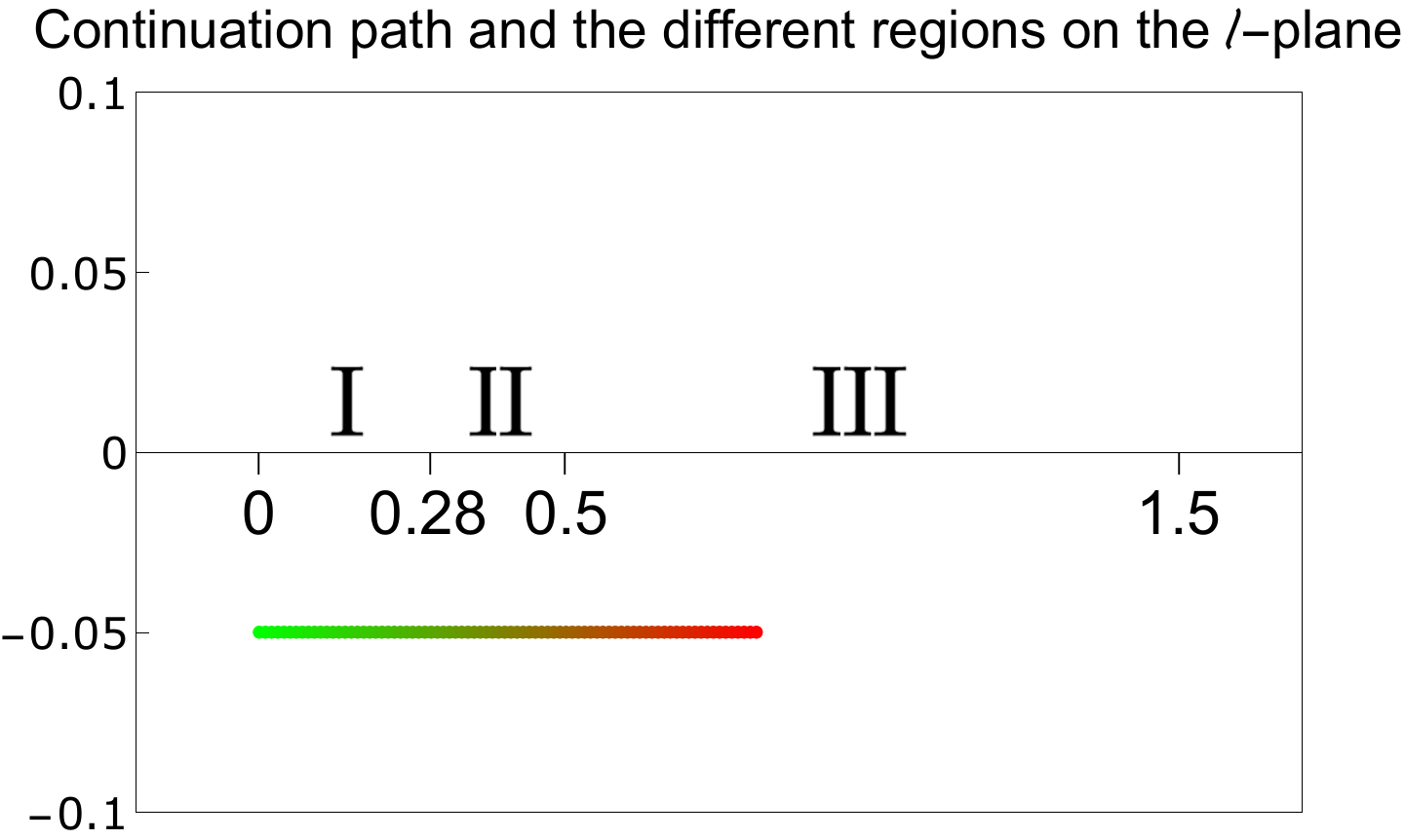}
	\caption{The evolution of the singularities of $\epsilon_1$ and $\hat\epsilon$ on the $\theta$-strip (left), located at $\theta=\A_j-i\pi/2$, $j=1,2$, and at $\theta=\tilde \A-i\pi/2$, as $\ell$ is analytically continued along a path below the real axis (right), from the unmodified region to the third modified region on the $\ell$-plane. In the left plot, the singularities of $\epsilon_1$ are shown as small circles, while those of $\hat{\epsilon}$ are shown as small crosses; the corresponding $\ell$ values are shown in the same colors in the right plot, where we have also labeled the modified regions as I, II, and III.}
	\label{fig:singularities2}
\end{figure}

Following a path on the complex $\ell$-plane, starting from $\ell\in(-1,0)$, going around $0$ to the positive real axis, zeros of $1+Y_1$ and $(1-e^{2\pi i \ell}\hat Y)(1-e^{-2\pi i \ell}\hat Y)$ cross the real line. Consequently the TBA equations receive residue corrections. We will refer to the neighborhoods of the successive segments of the positive real $\ell$-axis where more and more singularities cross the TBA contour as the ``$m$-th modified region", $m=1,2,3,\cdots$ (see right of Figure \ref{fig:singularities2}).

In the rest of this section we will work exclusively in the minimal chamber, specifically for the choice of parameters 
$$
(u_0, u_1) = (-3,1)
$$
in (\ref{eq:schrmeson}). The extension to other chambers by analytic continuation in $u_a$ (resulting in wall-crossing) is straightforward. Qualitative features of the modified regions around the positive real $\ell$-axis are not expected to be sensitive to the values of $u_a$.

\bigskip

\noindent{\bf The first modified region} occurs at $0<\ell \lesssim 0.28$.

Starting from $-1<\ell<0$, one can verify from the numerical solution to (\ref{eq:TBAnaiveell}) that both $1-e^{-2\pi i \ell} \hat Y(\theta)$ and $1-e^{2\pi i \ell} \hat Y(\theta)$ have zeros on the boundary of the $\theta$-strip, namely ${\rm Im}(\theta)=\pm\pi/2$. Approaching $\ell=0$ along the real $\ell$-axia, a pair of zeros, one for each of these functions, move toward $\theta=-\infty+ {i\pi\over 2}$ and $\theta=-\infty - {i\pi\over 2}$ respectively. Instead if we follow a small counterclockwise path on the complex $\ell$-plane around $\ell=0$, ${\rm Im}(\theta)$ increases for both zeros, so that one leaves the ${\rm Im}(\theta)<{\pi\over 2}$ strip while the other remains in the strip. We will refer to the latter zero as $\alpha_1(\ell)$. It is determined by the equation
\ie
\label{eq:consistencysingFirstmodified}
	2\pi i \ell = \hat \epsilon(\A_1(\ell))=\hat m e^{\alpha_1(\ell)} - \int\limits_{\mathbb{R}} {d\theta\over 2\pi} {\log(1+Y_1(\theta'))\over \cosh(\alpha_1(\ell)-\theta)}  .
\fe
As $\ell$ approaches the positive real axis from below, $\alpha_1(\ell)$ crosses the TBA integration contour, namely the real $\theta$-axis. Accounting for this, in the first modified region, the TBA equation for $\epsilon_1(\theta)$ (first equation of (\ref{eq:TBAnaiveell})) receives a correction term $\Delta \epsilon_1(\theta)$ on the RHS, given by
\ie\label{eq:MonodromyTBAell>0}
\Delta \epsilon_1(\theta) &= \oint\limits_{-\infty}^{\A_1(\ell)} {d\theta'\over 2\pi} \frac{\log\left((1-e^{2\pi i \ell}\hat Y(\theta'))(1-e^{-2\pi i \ell}\hat Y(\theta'))\right)}{\cosh(\theta-\theta')}
\\
&=-i\int\limits_{-\infty}^{\A_1(\ell)} \frac{ d\theta'}{\cosh(\theta-\theta')} = -\log \left(\frac{e^{\theta }+i e^{\A_1(\ell)}}{e^{\theta}-i e^{\A_1(\ell)}}\right) .
\fe
The integration contour in the first line is defined such that it encircles $(-\infty, \A_1]$ in counterclockwise direction as $\A_1(\ell)$ crosses the real axis. Collapsing the contour onto the branch cut of the logarithm results in the second line.

In conclusion, the TBA equations that are valid in the first modified region are
\ie
\label{eq:TBAsingfirstmodregion}
	\epsilon_1(\theta) &= m_1e^{\theta} - \int\limits_{\mathbb{R}} \frac{d\theta'}{2\pi} \frac{\log\left((1-e^{2\pi i \ell}\hat Y(\theta'))(1-e^{-2\pi i \ell}\hat Y(\theta'))\right)}{\cosh\left(\theta-\theta'\right)}-\log \left(\frac{e^{\theta}+i e^{\A_1(\ell)}}{e^{\theta}-i e^{\A_1(\ell)}}\right),\\
	\hat \epsilon(\theta) &= \hat m e^{\theta} - \int\limits_{\mathbb{R}} \frac{d\theta'}{2\pi} \frac{\log\left(1+Y_1(\theta')\right)}{\cosh\left(\theta-\theta'\right)} ,
\fe
with $\A_1(\ell)$ given by the solution of (\ref{eq:consistencysingFirstmodified}).
Note that
(\ref{eq:MonodromyTBAell>0}) is singular at $\theta=\A_1(\ell)\pm i\pi/2$, giving rise to a singularity of $\epsilon_1(\theta)$ itself at this location (see Figure \ref{fig:singularities2}). 
Furthermore, one finds that for real positive $\ell$, $\A_1(\ell)\in \mathbb{R}+{i\pi\over 2}$, and consequently the Y-functions take real values on the real $\theta$-axis. The iterative algorithm for solving these equations numerically is outlined in Appendix \ref{app:explicitNumerics}.

\bigskip

\noindent{\bf The second modified region} occurs at $0.28 \lesssim \ell \lesssim 0.5$.

As we analytically continue the solutions to the TBA equation along a path below the positive real $\ell$-axis, with increasing ${\rm Re}(\ell)$, the next singularity to cross the TBA integration contour comes from a zero of $1+Y_1(\theta)$. We denote its location on the $\theta$-strip by $\hat \alpha(\ell)$. After $\hat\A(\ell)$ crosses the integration contour, namely the real $\theta$-axis, we enter the second modified region, where (\ref{eq:TBAnaiveell}) receive the following corrections on the RHS,
\ie
	\label{eq:MonodromyTBAell>0_second}
	\Delta \epsilon_1(\theta) = -\log \left(\frac{e^{\theta}+i e^{\A_1(\ell)}}{e^{\theta}-i e^{\A_1(\ell)}}\right),~~~~
	\Delta \hat{\epsilon}(\theta) = -\log \left(\frac{e^{\theta}+i e^{\hat \alpha(\ell)}}{e^{\theta}-i e^{\hat \alpha(\ell)}}\right).
\fe
In other words, for this range of $\ell$, the TBA equations are modified to
\ie
	\label{eq:TBAsecregion}
	\epsilon_1(\theta) &= m_1e^{\theta} - \int\limits_{\mathbb{R}} \frac{d\theta'}{2\pi} \frac{\log\left((1-e^{2\pi i \ell}\hat Y(\theta'))(1-e^{-2\pi i \ell}\hat Y(\theta'))\right)}{\cosh\left(\theta-\theta'\right)}-\log \left(\frac{e^{\theta}+i e^{\A_1(\ell)}}{e^{\theta}-i e^{\A_1(\ell)}}\right),\\
	\hat \epsilon(\theta) &= \hat m e^{\theta} - 	\int\limits_{\mathbb{R}} \frac{d\theta'}{2\pi} \frac{\log\left(1+Y_1(\theta')\right)}{\cosh\left(\theta-\theta'\right)} - \log \left(\frac{e^{\theta}+i e^{\hat \alpha(\ell)}}{e^{\theta}-i e^{\hat \alpha(\ell)}}\right) ,
\fe
where $\A_1(\ell)$ and $\hat \A(\ell)$ are determined by the equations
\ie
\label{eq:secmod_thetas}
	2\pi i \ell &= \hat m e^{\A_1(\ell)} - \int\limits_{\mathbb{R}} \frac{\log(1+Y_1(\theta'))}{\cosh(\A_1(\ell)-\theta')}\frac{d\theta'}{2\pi} - \log \left(\frac{e^{\A_1(\ell)}+i e^{\hat \alpha(\ell)}}{e^{\A_1(\ell)}-i e^{\hat \alpha(\ell)}}\right) ,\\
	-\pi i &= m_1 e^{\hat \alpha(\ell)} - \int\limits_{\mathbb{R}} \frac{\log\left(\left(1-e^{2\pi i \ell}\hat Y(\theta')\right)\left(1-e^{-2\pi i \ell}\hat Y(\theta')\right)\right)}{\cosh(\hat \alpha(\ell)-\theta')}\frac{d\theta'}{2\pi} -\log \left(\frac{e^{\hat \alpha(\ell)}+i e^{\A_1(\ell)}}{e^{\hat \alpha(\ell)}-i e^{\A_1(\ell)}}\right) .
\fe

\bigskip

\noindent{\bf The third modified region} occurs at $0.5\lesssim \ell \lesssim 1.5$.

Increasing ${\rm Re}(\ell)$ further, below the positive real $\ell$-axis, a third singularity crosses the TBA contour as we enter the third modified region. This is the second singularity of $\log\left(1-e^{2\pi i \ell}\hat Y\right)$, whose location on the $\theta$-strip will be denoted $\A_2(\ell)$, determined by the equation $\hat \epsilon(\A_2) = 2\pi i (\ell-1)$. The resulting modified TBA equations are
\ie
	\epsilon_1(\theta) &= m_1e^{\theta} - \int\limits_{\mathbb{R}} \frac{d\theta'}{2\pi} \frac{\log\left((1-e^{2\pi i \ell}\hat Y(\theta'))(1-e^{-2\pi i \ell}\hat Y(\theta'))\right)}{\cosh\left(\theta-\theta'\right)}-\sum_{j=1}^{2}\log \left(\frac{e^{\theta}+i e^{\A_j(\ell)}}{e^{\theta}-i e^{\A_j(\ell)}}\right),\\
	\hat \epsilon(\theta) &= \hat m e^{\theta} - 	\int\limits_{\mathbb{R}} \frac{d\theta'}{2\pi} \frac{\log\left(1+Y_1(\theta')\right)}{\cosh\left(\theta-\theta'\right)} - \log \left(\frac{e^{\theta}+i e^{\hat \alpha(\ell)}}{e^{\theta}-i e^{\hat \alpha(\ell)}}\right) ,
\fe
where $\A_1$, $\A_2$, and $\hat\A$ are solved from
\ie{}
\label{eq:singLocsThirdMod}
&2\pi i (\ell+1-j) = \hat m e^{\A_j(\ell)} - \int\limits_{\mathbb{R}} \frac{d\theta'}{2\pi} \frac{\log\left(1+Y_1(\theta')\right)}{\cosh\left(\A_j(\ell)-\theta'\right)} - \log \left(\frac{e^{\A_j(\ell)}+i e^{\hat \alpha(\ell)}}{e^{\A_j(\ell)}-i e^{\hat \alpha(\ell)}}\right)  ,\\
& -\pi i = m_1 e^{\hat \A(\ell)} - \int\limits_{\mathbb{R}} \frac{d\theta'}{2\pi} \frac{\log\left((1-e^{2\pi i \ell}\hat Y(\theta'))(1-e^{-2\pi i \ell}\hat Y(\theta'))\right)}{\cosh\left(\hat \alpha(\ell)-\theta'\right)} - \sum_{j=1}^2\log \left(\frac{e^{\hat \alpha(\ell)}+i e^{\A_j(\ell)}}{e^{\hat \alpha(\ell)}-i e^{\A_j(\ell)}}\right)  .
\fe

\subsubsection{Numerical tests}
\label{sec:numTests}

As a consistency check, we can test the spectrum for the (analytically continued) non-relativistic meson 
obtained by numerically solving the TBA equations, described in the previous subsection
for positive real values of $\ell$ in the various modified regions,
against results obtained from Hamiltonian truncation (HT) method.

\begin{figure}[h!]
	\centering
	\includegraphics[width=0.45\linewidth]{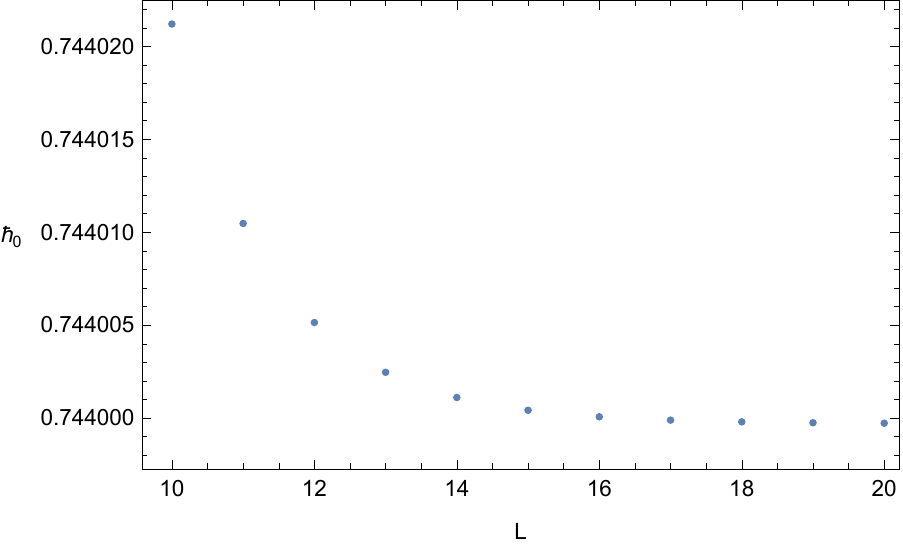}~~
	\includegraphics[width=0.45\linewidth]{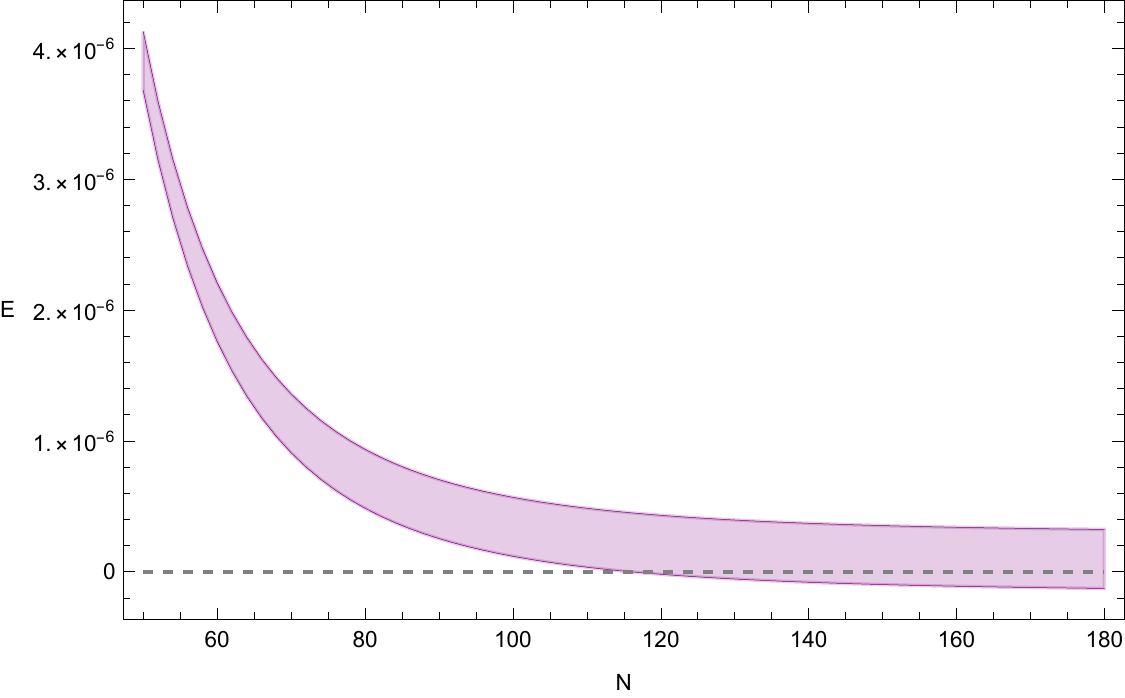}
	\caption{Test of TBA+EQC against Hamiltonian truncation (HT) for the ground state of the non-relativistic meson system, i.e. (\ref{eq:schrSing}) with $r=0$, at $(u_0, u_1) = (-3, 1)$ and $\ell=0.4$. Left: the value of $\hbar$ corresponding to the ground state, computed using the numerical solution to the TBA equations with a cutoff $L$ on the range of $\theta$-integration. The result converges to $\hbar =0.744$ in the large $L$ limit. Right: the purple band shows HT result for the ground state energy with $\hbar = 0.744 \pm 1.5 \times {10^{-7}}$, whereas the dotted line represents the expected value, namely zero, from the EQC at $\hbar=0.744$. 
}
	\label{fig:HT1}
\end{figure}

In the TBA+EQC method, to obtain the relation between $\hbar$ and the ground state energy for instance, we work at a given potential $V(x)$ whose constant (energy) term is arbitrarily chosen, and numerically determine the largest positive real value of $\hbar$ that solves the EQC. Then for comparison with HT method, we work at this value of $\hbar$ with the potential $V(x)$, and compute the ground state energy by diagonalizing the Hamiltonian on a finite dimensional subspace of the Hilbert space. The correct result for the ground state energy in this setting is, by definition, zero.

A sample of the comparison is shown in Figure \ref{fig:HT1}. Here we consider the Schr\"odinger problem (\ref{eq:schrSing}) with $r=0$ and $(u_0,u_1)=(-3,1)$, and $\ell=0.4$ which lies in the first modified region of section \ref{sec:NotwallCross}. To solve the TBA equations (\ref{eq:TBAsingfirstmodregion}), we adopt a cutoff $L$ on the range of the $\theta$-integration, and numerically extrapolate the resulting Voros spectrum to $L=\infty$. This method is highly accurate, as seen from the convergence in the left of Figure \ref{fig:HT1}. The Hamiltonian truncation approach, on the other hand, is subject to the systematic error in the extrapolation to infinite truncation level $N$. Comparison of the ground state energy computed using HT for a narrow window of $\hbar$ values centered around $\hbar_0=0.744$ (right of Figure \ref{fig:HT1}), against the ``correct" value of the ground state energy 0 at $\hbar=\hbar_0$, shows agreement well within 6 significant digits.

\begin{figure}[h!]
	\centering
	\includegraphics[width=0.45\linewidth]{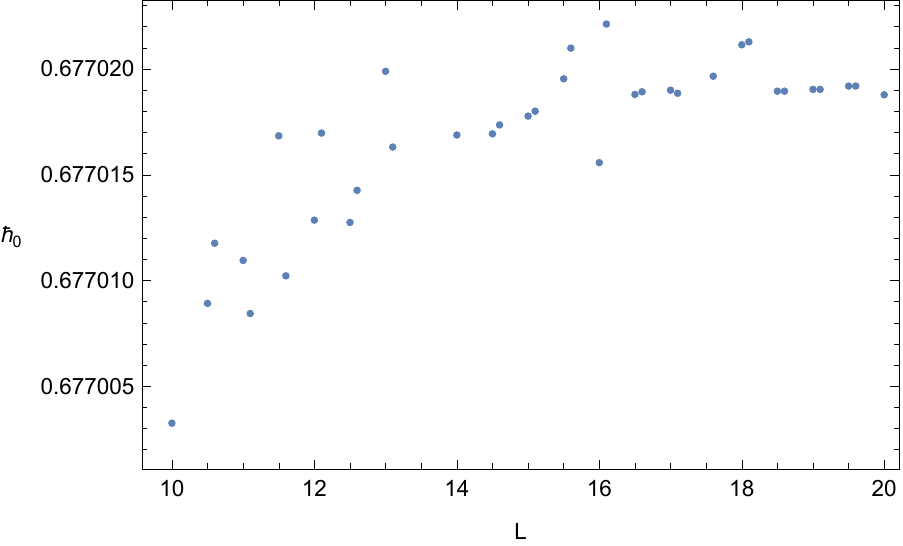}~~
	\includegraphics[width=0.45\linewidth]{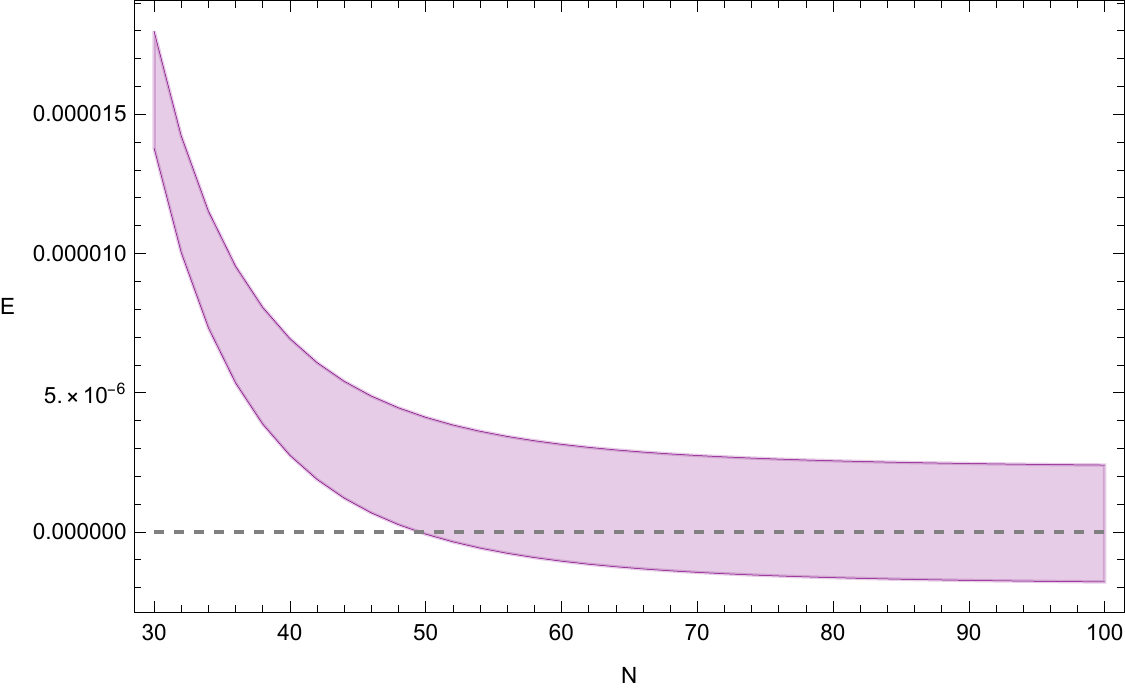}
	\caption{Test of TBA+EQC against HT in the same setting as Figure \ref{fig:HT1} but now for $\ell=0.6$ which lies in the third modified region. Left: a numerical instability is seen in the dependence on the cutoff $L$. Right: the purple band shows HT result for the ground state energy in the window $\hbar =0.677022 \pm 1.25 \times 10^{-6}$. 
	}
	\label{fig:numUnstable}
\end{figure}

We have repeated such tests for real positive values of $\ell$ in the second and third modified regions of section \ref{sec:NotwallCross}, as well as for other values of $(u_0, u_1)$, and found similar agreement. While we encountered some numerical instability in solving the TBA equation in the third modified region for $\ell<1$, the results are nonetheless in reasonably good agreement with HT, exemplified in Figure \ref{fig:numUnstable}.

\subsubsection{Analytic continuation in the complex $\ell$-plane}
\label{sec:AnContEll}

We now discuss some preliminary results concerning the analytic continuation of the Voros spectrum ($\hbar\equiv e^{-\theta}$ value) of the Schr\"odinger system (\ref{eq:schrSing}) in the complex $\ell$-plane, specializing to the non-relativistic meson case $r=0$, with the choice of $(u_0,u_1)=(-3,1)$ as in the previous subsection.

\begin{figure}[h!]
	\centering
	\includegraphics[width=0.43\linewidth]{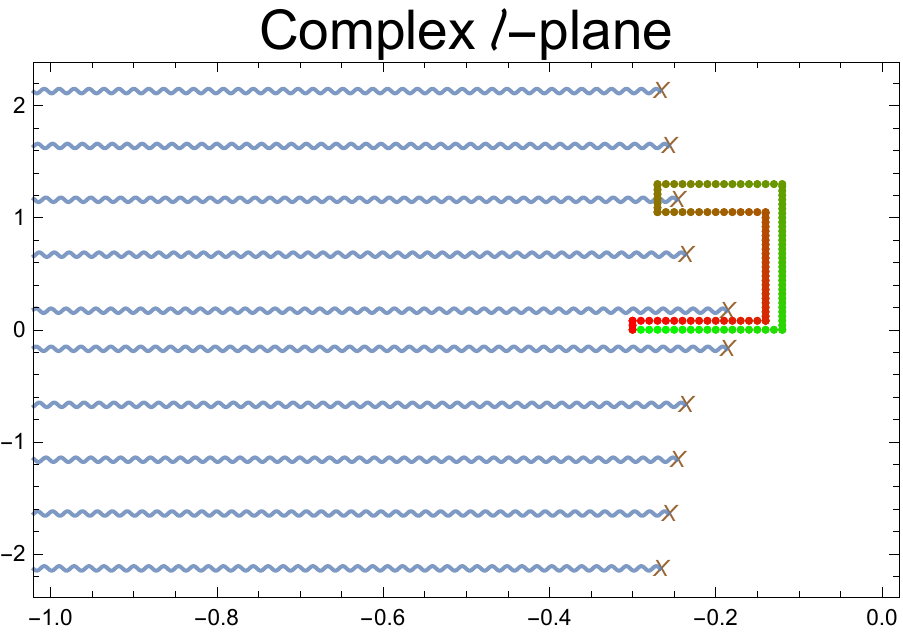}~~~
	\includegraphics[width=0.46\linewidth]{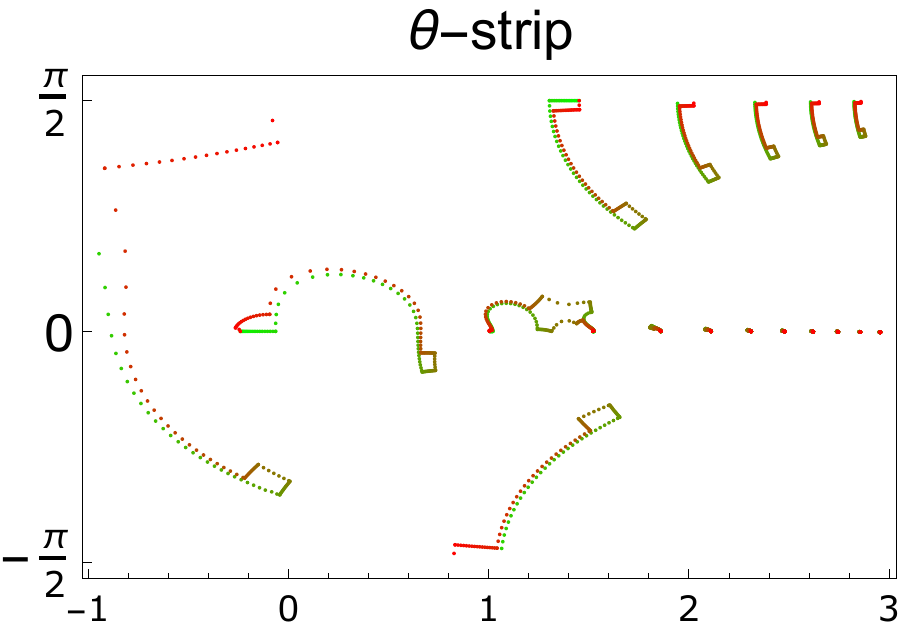} \\
	\vspace{.5cm}
	\includegraphics[width=0.43\linewidth]{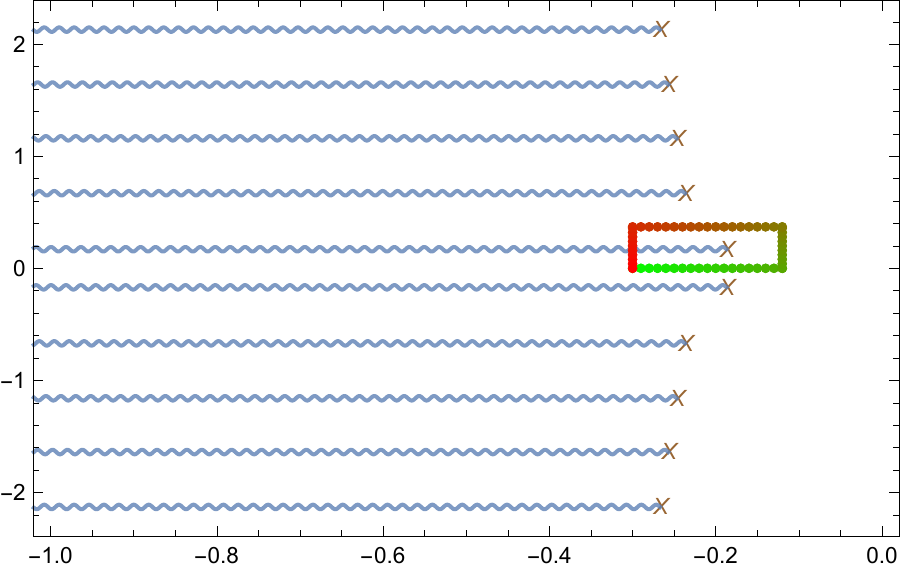}~~~
	\includegraphics[width=0.46\linewidth]{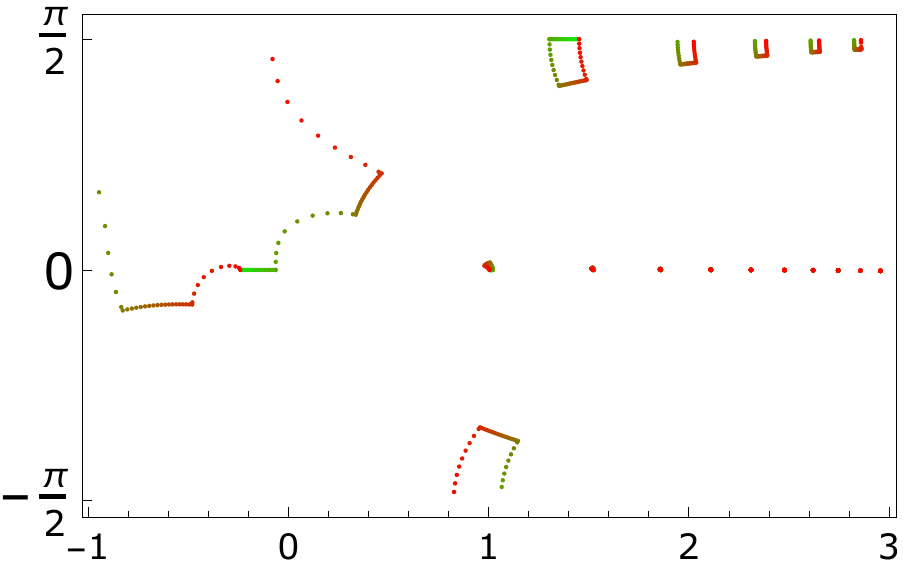}
	\caption{Analytic continuation of the Voros spectrum in the setting of section \ref{sec:nonrelMeson}. Left: the complex $\ell$-plane, with branch points marked by crosses, and a path of analytic continuation consisting of dots colored from green to red. Right: the solutions to the EQC on the $\theta$-strip, along the same path of analytic continuation shown in green to red. In the first row of plots, the monodromy around the third branch point exchanges the first and second excited states. In the second row, the monodromy around the first branch point exchanges the ground state with a complex (unphysical) solution to the EQC (see footnote \ref{footlabel}).}
	\label{fig:ellplane}
\end{figure}

For complex $\ell$ close to the ``unmodified region" $\ell\in [-1,0]$ on the real line, the spectrum is governed by the solutions to EQC, with quantum periods solved from the ``unmodified" TBA equation (\ref{eq:TBAnaiveell}). We numerically follow the analytically continued solution along a closed path on the complex $\ell$-plane, and observe that there are nontrivial monodromies around branch points on the $\ell$-plane under which a pair of solutions of the EQC are permuted. Two examples of such monodromies are shown in Figure \ref{fig:ellplane}.

The branch points on the complex $\ell$-plane come in complex conjugate pairs. We will order those on the upper half $\ell$-plane according to increasing ${\rm Im}(\ell)$, and refer to them as the first branch point, second branch points, etc. For $n\geq 2$, the monodromy around the $n$-th branch point on the principle sheet exchanges the $(n-1)$-th and $n$-th entries of the Voros spectrum (originally ordered along the real $\theta$-axis), as seen in the first row of Fig \ref{fig:ellplane}). On the other hand, the monodromy around the first branch point above the real $\ell$-axis exchanges the first entry of the Voros spectrum with an unphysical solution to the EQC, whose $\hbar$ value is close to being negative imaginary,\footnote{\label{footlabel}The solutions to the EQC with close-to-purely-imaginary $\hbar$ are not part of the physical Voros spectrum but they admit a simple physical interpretation: defining $\xi = e^{i\pi/2}\hbar$, we can write the Schroedinger equation (\ref{eq:schrSing}) with $r=0$ as $\left(-\xi^2 \partial_x^2 - x - u_{0} - u_1 x^{-1} + \xi^2 \ell(\ell+1)x^{-2}\right) \psi(x) = 0.$ The latter can be viewed a scattering problem that admits resonances corresponding to zeros of $W_{+0}$ close to the real $\xi$-axis (along similar lines of consideration of resonances in section \ref{sec:eqcnonsing}).
} 
 seen in the second row of Fig \ref{fig:ellplane}.

\section{Extraction of scattering phase}
\label{app:scatAmp}

In the previous sections we analyzed the quantization condition that determines the spectrum of resonances or bound states. For Schr\"odinger problems with an asymptotic region, it is also possible to use the Y-functions to extract the scattering phases in a similar fashion. The strategy is to fix the parameters of the potential including the energy, as well as the scattering phase $S(E)$, and find the values of $\hbar$ at which the desired scattering phase is realized. One can then numerically invert the relation to determine the phase as a function of $\hbar$. The procedure is analogous to the derivation of the EQC in section \ref{sec:EQC}.
Here we sketch a few examples.\footnote{These examples may be applied to analyzing the matrix quantum mechanics dual to deformations of 2D string theories \cite{Ginsparg:1993is, Martinec:2004td}, and in particular extracting the instanton expansion of the collective field scattering amplitudes that may be compared to worldsheet deformations along the lines of \cite{Balthazar:2019rnh, Balthazar:2019ypi}.}

\subsection{Polynomial potentials}

\paragraph{Polynomial potential of even $r$}
For a polynomial potential of odd degree of the form $V(x)=x^{r+1}+\cdots$, similarly to the discussion around (\ref{eqcreven}), the scattering wave function takes the form
\begin{equation}
	y_0 \propto S(E) y_{\frac{r+4}{2}} +  y_{\frac{r+2}{2}} .
\end{equation}
We can recast this relation as a ``quantization condition"
\begin{equation}
	S(E) W_{0,\frac{r+4}{2}} +  W_{0,\frac{r+2}{2}} = 0 ,
\end{equation}
which can then be expressed in terms of the Y-functions in a manner similar to that of section \ref{sec:eqcnonsing}.

\paragraph{Inverted polynomial potential of odd $\mathbf{r}$} The Schr\"odinger problem with a polynomial potential of the form $V(x)=-x^{r+1}+\cdots$ has two asymptotic regions $x\to \pm\infty$.
The derivation of the TBA equations that govern the quantum periods is very similar to the one in section \ref{sec:ExactPoly} with two minor modifications. First, the WKB cycles corresponding to classically allowed and forbidden regions are exchanged. Second, the basis of wave functions $y_k$ are now defined with the prescribed asymptotic behavior on the wedges $\widetilde{\cal S}_k$,
\ie
	\label{eq:sectorsINVE}
	 \widetilde{\cal S}_k = \Bigg\{ x \in \mathbb{C}  : \left|\arg(x)-\frac{2\pi (k-1/2)}{r+3}\right| < \frac{\pi}{r+3}\Bigg\} .
\fe
A scattering state with incoming flux from the right has the wave function
\begin{equation}
	y_{0} \propto S_{T}(E) y_{\frac{r+5}{2}} +  S_R(E) y_{1},
\end{equation}
where $S_{R}$ and $S_{T}$ are the reflection and transmission amplitudes respectively. The EQC then determines $\hbar$ as a function of $(S_R,S_T,E)$.

\subsection{Potentials with a regular singularity}

Analogously to (\ref{eq:schrSing}), we can consider the Schr\"odinger problem with a regular singularity and a potential unbounded from below, 
\begin{equation} \label{eq:schrSingINVE}
	\left(-\hbar^2 \partial_x^2 - x^{r+1} + \sum_{a=0}^{r+1} u_{a} x^{r-a} + \frac{\hbar^2 \ell(\ell+1)}{x^2}\right) \psi(x) = 0 .
\end{equation}
The basis wave functions $y_k$ can again be defined by their asymptotics on the wedges $\widetilde{\cal S}_k$ as in (\ref{eq:sectorsINVE}). The scattering state of interest takes the form
\begin{equation}
	\psi_\pm \propto S(E) y_{1} +  y_{0} ,
\end{equation}
where the subscript $\pm$ specifies the behavior of the wave function near $x=0$, as in (\ref{ynearorign}).
By manipulations similar to that of section \ref{sec:exact_sing}, we have
\begin{equation}
	S(E) y_{r+4} + y_{r+3} = \omega^{(r+3)/2}e^{\pm 2\pi i \ell}\left(S(E) y_{1} +  y_{0}\right) ,
\end{equation}
which may be equivalently expressed in terms of the Wronskian conditions
\ie\label{quansca}
&	W_{r+4, r+3} = \omega^{(r+3)/2}e^{\pm 2\pi i \ell}\left(S(E) W_{{r+4},1} +  W_{{r+4}, 0}\right)  ,\\
&	S(E) W_{r+3,r+4}  = \omega^{(r+3)/2}e^{\pm 2\pi i \ell}\left(S(E) W_{r+3,1} + W_{r+3,0}\right)  .
\fe
We can further eliminate $W_{r+4,0}$ using the Plucker relation $W_{{r+4}, 0}W_{r+3, 1} = W_{r+4, 1} W_{r+3,0} - W_{0, 1} W_{r+3, r+4}$. In the case of even $r$, $W_{r+4,1}$ and $W_{r+3,0}$ can be expressed in terms of the $Y_s$ functions with odd $s$, whereas $W_{r+3,1}$ can be expressed in terms of $\hat Y$. Hence, the quantization condition (\ref{quansca}) can be written in terms of the Y-functions entirely, as desired. The case of odd $r$ can be treated similarly.

\section{Discussion}
\label{sec:discussion}

Let us summarize the TBA+EQC approach to Schr\"odinger problems discussed in this paper. At a given energy, the TBA equations determine the quantum periods as functions of $\hbar$. The EQC based on the quantum periods then determines the values of $\hbar$ for bound states or resonances at the given energy (Voros spectrum). 

For Schr\"odinger problems with an arbitrary polynomial potential, the TBA equations are by now well known \cite{Balian:1978et, AIHPA_1983__39_3_211_0, DelErHerv_ExactSemi, AIHPA_1999__71_1_1_0, Voros:1999bz, Ito:2018eon,Emery:2020qqu}, as reviewed in section \ref{sec:ExactPoly}. In the case of potential with a regular singularity, which may be viewed as a spherically symmetric system with polynomial potential in a sector of angular momentum $\ell$, the relevant TBA equations are obtained in \cite{Ito:2020ueb} for the range $-1<\ell<0$. We pointed out that while the TBA equations of \cite{Ito:2020ueb} do not hold for general angular momentum $\ell$, the correct ``modified" TBA equations for $\ell>0$ can be obtained by careful analytic continuation.

As for the EQC, in the polynomial potential case a derivation is in principle available by solving the ``connection problem" \cite{Voros:1999bz, PhysRevLett.55.2523}. In this paper, we gave a streamlined derivation through elementary manipulation of Wronskian relations, which has not previously appeared in the literature to the best of our knowledge. More importantly, we extended this derivation of the EQC to the case of potential with a regular singularity, completing the system of equations required to solve the spectral problem. The latter is shown to pass numerical checks against results from Hamiltonian truncation method.

The TBA+EQC method is applied to study the analytic continuation of the non-relativistic meson system in the angular momentum $\ell$ in section \ref{sec:AnContEll}. We found that the Voros spectrum has a set of branch points on the complex $\ell$-plane, around which monodromies occur. A more extensive investigation of the modified TBA equation on the entirety of the complex $\ell$-plane, as well as the analytic property of the spectrum thereof, is left to future work.

One important question concerning the exact quantization method is how to generalize it to quantum mechanical systems with more than one degree of freedom. This is conceivable at least for quantum systems whose classical limit correspond to integrable Hamiltonian systems (in a given energy range), where one may hope to construct EQCs from quantum analogs of action variables. Another question, which served as the original motivation of this work, is whether the non-perturbative handle of the analytic continuation of quantum mechanical systems in coupling parameters considered here can be extended to quantum field theories \cite{Fonseca:2001dc}. We hope to return to these questions in the future.

\section*{Acknowledgements} 

This work is supported in part by a Simons Investigator Award from the Simons Foundation, by the Simons Collaboration Grant on the Non-Perturbative Bootstrap, and by DOE grant DE-SC0007870. The numerical computations in this work are performed on the Odyssey cluster supported by the FAS Division of Science, Research Computing Group at Harvard University.

\appendix

\section{Wall-crossing of the TBA equation}
\label{sec:wallCrossing}

As we vary the parameters (including energy) in the potential of the Schr\"odinger system, sometimes one encounters wall-crossing phenomena in which the classical limit of the Y-functions jump (as do the WKB cycles), whereas in fact at finite $\hbar$ the Y-functions remain analytic in the parameters and the EQC remain invariant when expressed in terms of the same set of Y-functions. 

Of particular physical interest are complex paths in the parameter space that start and end with real parameters. As we complexify the potential $V(x)$, starting from the vicinity of the domain (\ref{minorder}), the WKB periods $m_a\equiv |m_a|e^{i\phi_a}$ become complex. Let us define
\ie\label{epsilonelldef}
\tilde \epsilon_a(\theta) \equiv \epsilon_a(\theta-i\phi_a) , ~~~~ \tilde L_a(\theta) \equiv \log(1+e^{-\tilde\epsilon_a(\theta)}) = L_a(\theta-i\phi_a) ,
\fe
and
\begin{equation}
K_{a,b}(\theta) \equiv  K(\theta-i\phi_a+i\phi_b),
\end{equation}
so that the TBA equation (\ref{tbaeqns}) can be put in the form
\begin{equation}\label{newmtba}
\tilde \epsilon_a = |m_a|e^{\theta} - K_{a,a+1}\star \tilde L_{a+1} - K_{a,a-1}\star \tilde L_{a-1},
\end{equation}
provided $|\phi_{a}-\phi_{a\pm 1}|<{\pi\over 2}$. When the latter condition is violated, a pole of the kernel $K$ crosses the $\theta$-integration contour in the TBA equation, which must be compensated by a residue contribution to maintain the analyticity of the solution. This is the wall-crossing phenomenon from the perspective of the TBA equation.

In \cite{Ito:2018eon} the cubic potential case $r=2$ is analyzed explicitly, and its generalization is straightforward. Here we discuss an example of the wall-crossing for the quartic potential ($r=3$) considered in section \ref{sec:quadcase}. Specifically, we consider the wall-crossing that occurs at 
\ie
\phi_{2}-\phi_{1}={\pi\over 2},
\fe
after which the TBA equation (\ref{newmtba}) is modified to
\ie\label{modntba}
\tilde \epsilon_1(\theta) &= |m_1|e^{\theta} - K_{1,2}\star \tilde L_{2} - L_{2}\left(\theta-i\phi_1-\frac{i\pi}{2}+i0\right) ,\\
\tilde \epsilon_2(\theta) &= |m_2|e^{\theta} - K_{2,3}\star \tilde L_{3} - K_{2,1}\star \tilde L_{1} - L_{1}\left(\theta-i\phi_2+\frac{i\pi}{2}-i0\right) ,\\
\tilde \epsilon_3(\theta) &= |m_3|e^{\theta} - K_{3,2}\star \tilde L_{2}  .
\fe
This is one form of the wall-crossed TBA equation that has appeared in the literature (e.g. in \cite{Alday:2010vh}).
To close this set of equations requires taking into account the analytic continuation that relates the $L_a$ functions appearing on the RHS to $\tilde\epsilon_a(\theta)$, which is inconvenient for numerical solution by iteration.

It is possible to turn the equations  (\ref{modntba}) into a more standard TBA form by defining a new set of Y-functions $Y^n_a$,
\ie\label{ynewrely}
& Y_1^n (\theta) \equiv \frac{Y_1(\theta)}{1+Y_2(\theta-i\pi/2)} ,~~~~ Y_2^n (\theta) \equiv \frac{Y_2(\theta)}{1+Y_1(\theta+i\pi/2)} ,\\
& Y_3^n (\theta) \equiv Y_3(\theta)  ,~~~~ Y_{12}^n (\theta) \equiv \frac{Y_1(\theta)Y_2(\theta-i\pi/2)}{1+Y_1(\theta)+Y_2(\theta-i\pi/2)},
\fe
and the corresponding $\epsilon_a^n$ by $Y_a^n(\theta) \equiv \exp\left[-\epsilon_a^n(\theta)\right]$. 
We will define $\tilde\epsilon_a^n$ and $\tilde L_a^n$ by shifting the imaginary part of the argument as in (\ref{epsilonelldef}). With these, (\ref{modntba}) is now recast in the form
\ie
\tilde \epsilon^n_1 &= |m_1|e^{\theta} - K_{1,2}\star \tilde L^n_{2} - K^+_{1,12}\star \tilde L^n_{12},
\\
\tilde \epsilon^n_2 &= |m_2|e^{\theta} - K_{2,1}\star \tilde L^n_{1} - K_{2,3}\star \tilde L^n_{3} - K_{2,12}\star \tilde L^n_{12},
\\
\tilde \epsilon^n_3 &= |m_3|e^{\theta} - K_{3,2}\star \tilde L^n_{2} - K^+_{3,12}\star \tilde L^n_{12},
\\
\tilde \epsilon^n_{12} &= |m_{12}|e^{\theta} - K^-_{12,1}\star \tilde L^n_{1}  - K^-_{12,3}\star \tilde L^n_{3} - K_{12,2}\star \tilde L^n_{2},
\fe
where $m_{12}\equiv m_1-im_2$ and $K_{a,b}^\pm(\theta)\equiv K_{a,b}(\theta \pm i\pi/2)$.

Note that the EQC found in section \ref{sec:EQC}, as a set of constraints on the Y-functions, is invariant under wall-crossing. For the purpose of finding numerical solutions, it is useful to re-express the constraints in terms of the new Y-functions in the relevant chamber using relations analogous to (\ref{ynewrely}).

\section{Some numerical details}
\label{app:explicitNumerics}

We outline our numerical algorithm for solving the modified TBA systems appearing in section \ref{sec:NotwallCross}. Schematically, an integral equation of the form $\Phi = F[\Phi]$ may be solved by iteration, starting with a seed $\Phi^{(0)}$ and set $\Phi^{(n+1)} = F[\Phi^{(n)}]$, such that $\Phi^{(n)}$ converges to the desired solution as $n\to \infty$. In practice, it is useful to set $\Phi^{(n+1)} = aF[\Phi^{(n)}] + (1-a)\Phi^{(n)}$ with a suitable choice of the parameter $a$ for numerical stability. The typical number of iterations we have used is $\sim 10^4$, which is more than sufficient for the desired accuracy.

The integrations appearing in the iterative solution of $\hat \epsilon$ and $\epsilon_1$ are evaluated using discrete fast Fourier transform. The integration contour in $\theta$-strip is discretized using a grid of $N_{\rm max}$ points up to a cutoff $L$. In most cases we sample $N_{\rm max}$ up to $3\times 10^4$, 
and then extrapolate the result to $N_{\rm max}=\infty$ at a fixed $L$. $L$ is typically taken between $10$ and $20$, without the need for further numerical extrapolation (see e.g. figure \ref{fig:HT1}). 

In the first modified region, the TBA equations are given by (\ref{eq:TBAsingfirstmodregion}) together with (\ref{eq:consistencysingFirstmodified}). It is useful to work with $\theta\in\mathbb{R}+i\varphi$ for some nonzero $\varphi$, and deform the $\theta$-integration contours accordingly so as to avoid the singularities (related to $\alpha_1(\ell)$) of the integrand on the RHS of (\ref{eq:consistencysingFirstmodified}). The numerical algorithm is as follows.
\begin{enumerate}
	\item{Set $\varphi={\pi\over 4}$, and set the seed values of $\hat\epsilon$, $\epsilon_1$ to be $\hat m e^{\theta}, m_1 e^{\theta}$. The seed value of $\A_1(\ell)$ is taken to be $a+i{\pi\over 2}$ for some arbitrary real $a$.}
	\item{Iterate the equation (\ref{eq:consistencysingFirstmodified}) for $\alpha_1(\ell)$ until numerical convergence.}
	\item{Iterate the equations (\ref{eq:TBAsingfirstmodregion}) for $\hat\epsilon$, $\epsilon_1$ once.}
	\item{Repeat steps 2 and 3 a large number of times.}
	\item{The numerical value of ${\rm Im}\, \alpha_1(\ell)$ should be close to ${\pi\over 2}$, and we now set it to be ${\pi\over 2}$. Plugging the result for $\A_1(\ell)$ into (\ref{eq:TBAsingfirstmodregion}), we numerically solve $\epsilon_1,\hat \epsilon$ on the real $\theta$-axis (i.e. setting $\varphi=0$) by a large number of iterations. After each iteration, the imaginary part of $\hat\epsilon$, $\epsilon_1$ should be numerically insignificant, and will be set to $0$ for numerical stability.}
\end{enumerate}
For $2\times 10^4$ total iterations, the computation for each value of $N_{\rm max}$ and $L$ typically takes a few hours on one core with $1.5$ GB memory at the Harvard research computing cluster.

In the second modified region, the TBA equations are given by (\ref{eq:TBAsecregion}) and (\ref{eq:secmod_thetas}). 
The algorithm is similar to the first modified region but now we must also solve $\hat \A(\ell)$ by iteration of (\ref{eq:secmod_thetas}).

In the third modified region, for $\ell>1$ the numerical algorithm is similar to above. For $\ell<1$, we use local minimization of the equation rather than iteration to solve for $\A_2(\ell)$, as the latter method can lead to misidentification of the location of $\A_2(\ell)$ on the $\theta$-strip.

\bigskip

The numerical implementation of Hamiltonian truncation method for the non-relativistic meson system of section \ref{sec:numTests} involves the choice of an IR cutoff $L$ and the truncation level $N$. The potential is decomposed as $V=V_0+V_1$, where $V_0$ is the angular momentum term that will be included in the Hamiltonian $H_0$, and
\ie
	V_1 = x + u_1 + \frac{u_2}{x}
\fe
is viewed as the deformation. The wave functions for an eigenbasis of $H_0$ with a hard wall at $x=L$ are
\ie
	\Psi_n(x) \propto \mathcal{N}\sqrt{x} J_{(1+2\ell)/2}\left( \frac{Z^n_{(1+2\ell)/2}\, x}{L} \right) ,
\fe
where $Z^n_j$ is the $n$-th zero of the $j$-th Bessel function. 
We then numerically evaluate the matrix elements of $V_1$ with respect to this basis, up to $n\leq N$, and diagonalize the corresponding truncated full Hamiltonian matrix.

For the purpose of testing against EQC+TBA results, we compute the spectrum using the HT method with $L=10$ and truncation levels $N$ between $30$ and $1140$, and extrapolate the results to $N=\infty$. On one core with 1.5 GB memory, the runtime is about $20$ seconds for $N=30$, and $60$ hours for $N=1140$.

\bibliographystyle{JHEP}
\bibliography{QMvac}

\end{document}